\documentclass[longauth]{aa} 
\usepackage[normalem]{ulem}
\usepackage{graphicx}
\usepackage{txfonts}
\usepackage{siunitx}
\usepackage{longtable} 
\usepackage{lscape}
\usepackage{rotating}
\usepackage{natbib}
\usepackage[normalem]{ulem}
\DeclareSIUnit\gauss{G}
\begin{document}
\renewcommand{\arraystretch}{1.0}
\title{The CARMENES search for exoplanets around M dwarfs}
\subtitle{A long-period planet around GJ\,1151\\
measured with CARMENES and HARPS-N data\thanks{Tables C.1 and C.2 are only available in electronic form at the CDS via anonymous ftp to \url{cdsarc.cds.unistra.fr (130.79.128.5)}
or via \url{https://cdsarc.cds.unistra.fr/cgi-bin/qcat?J/A+A/}}}
\author{
J.~Blanco-Pozo\inst{1,2}
\and M.\,Perger\inst{1,2} 
\and M.\,Damasso\inst{3} 
\and G.~Anglada Escudé\inst{1,2} 
\and I.~Ribas\inst{1,2}
%
\and D.~Baroch\inst{1,2} 
\and J.\,A.~Caballero\inst{4} 
\and C.\,Cifuentes\inst{4} 
\and S.\,V.~Jeffers\inst{5} 
\and M.~Lafarga\inst{1,2,6} 
\and A.~Kaminski\inst{7} 
\and S.~Kaur\inst{1,2}  
\and E.~Nagel\inst{8,9,10} 
\and V.~Perdelwitz\inst{11} 
\and M.\,Pérez-Torres\inst{12,13} 
\and A.\,Sozzetti\inst{3} 
\and D.~Viganò\inst{1,2} 
%
\and P.\,J.~Amado\inst{12} 
\and G.~Andreuzzi\inst{14,15} 
\and V.\,J.\,S.\,B\'ejar\inst{16,17}
\and E.\,L.\,Brown\inst{18} 
\and F.~Del\,Sordo\inst{1,2,19} 
\and S.~Dreizler\inst{8} 
\and D.~Galad\'i-Enr\'iquez\inst{20} 
\and A.\,P.~Hatzes\inst{10} 
\and M.~K{\"u}rster\inst{21} 
\and A.\,F.\,Lanza\inst{19}  
\and A.\,Melis\inst{22}  
\and E.\,Molinari\inst{22} 
\and D.~Montes\inst{23} 
\and M.\,Murgia\inst{22} 
\and E.~Pall\'e\inst{16,17} 
\and L.~Peña-Moñino\inst{12} 
\and D.\,Perrodin\inst{22} 
\and M.\,Pilia\inst{22} 
\and E.\,Poretti\inst{14} 
\and A.~Quirrenbach\inst{9} 
\and A.~Reiners\inst{8} 
\and A.~Schweitzer\inst{10} 
\and M.~R.~Zapatero Osorio\inst{24} 
\and M.~Zechmeister\inst{8} 
}

\institute{
\inst{1}Institut de Ci\`encies de l'Espai (ICE, CSIC), Campus UAB, Carrer de Can Magrans s/n, 08193 Bellaterra, Spain\\
\inst{2}Institut d'Estudis Espacials de Catalunya (IEEC), 08034 Barcelona, Spain\\
\email{blanco@ice.csic.es} \\
\inst{3}INAF - Osservatorio Astrofisico di Torino, Via Osservatorio 20, 10025 Pino Torinese, Italy\\
\inst{4}Centro de Astrobiolog\'{\i}a (CAB), CSIC-INTA, ESAC Campus, Camino bajo del castillo s/n, 28692 Villanueva de la Ca\~nada, Madrid, Spain\\
\inst{5}Max Planck Institut f\"ur Sonnensystemforschung (MPS), Justus-von-Liebig-Weg 3, 37077 G\"ottingen, Germany\\
\inst{6}Department of Physics, University of Warwick, Gibbet Hill Road, Coventry CV4 7AL, UK\\
\inst{7}Landessternwarte, Zentrum f\"ur Astronomie der Universit\"at Heidelberg, K\"onigstuhl 12, 69117 Heidelberg, Germany\\
\inst{8}Institut f\"ur Astrophysik, Georg-August-Universit\"at G\"ottingen, Friedrich-Hund-Platz 1, 37077 G\"ottingen, Germany\\
\inst{9}Hamburger Sternwarte, Gojenbergsweg 112, 21029 Hamburg, Germany\\
\inst{10}Th\"uringen Landessternwarte Tautenburg, Sternwarte 5, 07778 Tautenburg, Germany\\
\inst{11}Kimmel Fellow, Department of Earth \& Planetary Sciences, Weizmann Institute of Science, Rehovot, Israel\\
\inst{12}Instituto de Astrof\'{\i}sica de Andaluc\'{\i}a (IAA-CSIC), Glorieta de la Astronom\'{\i}a s/n, 18008 Granada, Spain\\
\inst{13}School of Sciences, European University Cyprus, Diogenes street, Engomi, 1516 Nicosia, Cyprus\\
\inst{14}Fundaci\'on Galileo Galilei-INAF, Rambla Jos\'e Ana Fernandez P\'erez 7, 38712 Bre\~na Baja, TF, Spain\\
\inst{15}INAF - Osservatorio Astronomico di Roma, Via Frascati 33, 00078 Monte Porzio Catone, Italy\\
\inst{16}Instituto de Astrof\'{\i}sica de Canarias (IAC), 38205 La Laguna, Tenerife, Spain\\
\inst{17}Departamento de Astrof\'{\i}sica, Universidad de La Laguna (ULL), 38206, La Laguna, Tenerife, Spain\\
\inst{18}University of Southern Queensland, Centre for Astrophysics, Toowoomba, QLD, 4350, Australia\\
\inst{19}INAF - Osservatorio Astrofisico di Catania, Via Santa Sofia 78, 95123 Catania, Italy\\
\inst{20}Centro Astron\'omico Hispano en Andaluc\'{\i}a (CAHA), Observatorio de Calar Alto, Sierra de los Filabres, 04550 G\'ergal, Spain\\
\inst{21}Max Planck Institut f\"ur Astronomie (MPIA), K\"onigstuhl 17, 69117 Heidelberg, Germany\\
\inst{22}INAF - Osservatorio Astronomico di Cagliari Via della Scienza 5, 09047 Selargius, Italy\\
\inst{23}Departamento de F\'{\i}sica de la Tierra y Astrof\'{\i}sica \& IPARCOS-UCM (Instituto de F\'{\i}sica de Part\'{\i}culas y del Cosmos de la UCM), Facultad de Ciencias F\'{\i}sicas, Universidad Complutense de Madrid, 28040 Madrid, Spain\\
\inst{24}Centro de Astrobiolog\'{\i}a (CAB), CSIC-INTA, Carretera de Ajalvir km 4, 28850 Torrej\'on de Ardoz, Madrid, Spain\\
}
\date{Received September 26, 2022; accepted December 24, 2022}
\abstract
{Detecting a planetary companion in a short-period orbit through radio emission from the interaction with its host star is a new prospect in exoplanet science. Recently, a tantalising signal was found  close to the low-mass stellar system GJ\,1151 using LOFAR observations.}
{We studied spectroscopic time-series data of GJ\,1151 in order to search for planetary companions, investigate possible signatures of stellar magnetic activity, and to find possible explanations for the radio signal.}
{We used the combined radial velocities measured from spectra acquired with the CARMENES, HARPS-N, and HPF instruments, extracted activity indices from those spectra in order to mitigate the impact of stellar magnetic activity on the data, and performed a detailed analysis of {\it Gaia} astrometry and all available photometric time series coming from the MEarth and ASAS-SN surveys.}
{We found a M$>$10.6\,M$_{\oplus}$ companion to GJ\,1151 in a 390d orbit at a separation of 0.57\,au. Evidence for a second modulation is also present; this could be due to long-term magnetic variability or a second (substellar) companion. The star shows episodes of elevated magnetic activity, one of which could be linked to the observed LOFAR radio emission. We show that it is highly unlikely that the detected GJ\,1151\,b, or any additional outer companion is the source of the detected signal. We cannot firmly rule out the suggested explanation of an undetected short-period planet that could be related
to the radio emission,  as we establish an upper limit of 1.2\,M$_{\oplus}$ for the minimum mass.}
{}
\maketitle
%
\section{Introduction} \label{sec:intro}

The field of exoplanets has evolved over the last decades, from detecting massive gas giants \citep{1995Natur.378..355M}, to ground-based searches around samples of nearby low-mass stars with high-resolution spectrographs \citep[e.g. HARPS, HARPS-N, CARMENES, ESPRESSO;][]{1994SPIE.2198..362V, 2003Msngr.114...20M, 2009Msngr.136...39K, 2010SPIE.7735E..0FP, 2012SPIE.8446E..1VC, 2013A&A...549A.109B, 2015A&A...581A..38C, 2016A&A...593A.117A, 2017AAS...22940307J, 2018A&A...609L...5R, 2020SPIE11447E..3CQ} and space-based photometers searching for transits in large-scale surveys such as CoRoT \citep{2006cosp...36.3749B}, {\it Kepler} \citep{2011ApJ...736...19B} or TESS \citep[Transiting Exoplanet Survey Satellite;][]{2015JATIS...1a4003R}, to the analysis of planetary atmospheres \citep{2007Natur.448..169T, 2008A&A...487..357S, 2019A&A...632A..69Y, 2019AJ....158...91S, 2022A&A...657A...6C}. Nowadays, additional techniques to reveal and analyse exoplanets are becoming increasingly interesting \citep{2018exha.book.....P}. One of these new methods is the detection of low-frequency radio emission \citep{1998JGR...10320159Z, 2015Natur.523..568H}.

\citet{2020NatAs...4..577V} announced the detection of radio emission with the LOFAR instrument \citep[LOw Frequency ARray;][]{2013A&A...556A...2V} at the position of the M4.5-dwarf GJ\,1151. The instrument is a European radio interferometric antenna network mostly based in the Netherlands and operating at frequencies from 10 to 240\,MHz or wavelengths from 1.2 to 30\,m. The high degree of circular polarisation ($64\pm6\,\%$), the relatively long duration ($>$8\,h) of the emission, and the low activity of the star makes the signal compatible with stellar auroral radio emission triggered by a substellar companion.  \citet{2020NatAs...4..577V} concluded that the observed radio emission could not originate from the slow rotation of GJ\,1151, but could be powered by the sub-Alfvénic interaction of the star with a planet comparable to the mass of the Earth with an orbital period of 1 to 5\,days.

This finding prompted spectroscopic follow-up observations of this star in order to search for a planet able to produce the detected radio emission. \citet{2020ApJ...890L..19P} used 19 radial velocity (RV) measurements from HARPS-N \citep[High Accuracy Radial velocity Planet Searcher in the Northern hemisphere;][]{2012SPIE.8446E..1VC}  found no significant signal. The authors set an upper limit on the minimum mass of the possible companion at $M\sin{i} < 5.6\,{\rm M}_\oplus$. \citet{2021ApJ...919L...9M} added 24 RVs extracted and binned from 50 near-infrared (NIR) spectra obtained with the HPF \citep[Habitable-zone Planet Finder;][]{2012SPIE.8446E..1SM} instrument. These authors claimed the detection of a significant signal caused by a possible planet with minimum mass $M\sin{i}=2.5\pm0.5\,{\rm M}_\oplus$, a period of 2.02\,d, and an RV amplitude of 4.1\,m\,s$^{-1}$. However, \citet{2021A&A...649L..12P} reported the absence of such a signal after combining the same data from HPF and HARPS-N with an additional 70 CARMENES \citep[Calar Alto high-Resolution search for M dwarfs with Exo-earths with Near-infrared and optical Échelle Spectrographs;][]{2020SPIE11447E..3CQ} RVs extracted from the visible channel of the instrument. The authors calculated detection limits for the possible low-mass planet in 1d and 5d orbits of 0.7\,M$_\oplus$ and 1.2\,M$_\oplus$, respectively. Additionally, they found evidence for a signal at an orbital period of $>$300\,d, which we investigate further in the present study. GJ\,1151 is an M4.5 dwarf star at a distance of 8\,pc from the Sun with a low magnetic activity level. We list the relevant parameters of the star in Table\,\ref{table:GJ1151data}.

\begin{table}
\caption{Stellar parameters for GJ\,1151 \citep{1979A&AS...38..423G}, G\,122-49 \citep{1971lpms.book.....G}, or Karmn\,J11509+483 \citep{2013hsa7.conf..645C}.} \label{table:GJ1151data}
\centering  
\begin{tabular}{l c c}     
\hline\hline \noalign{\smallskip} 
Parameter       &        Value  &  Ref. \\ 
\noalign{\smallskip} \hline \noalign{\smallskip}
Sp. type & M4.5\,V & (1) \\
$\alpha$ (J2016) & 11h 50m 55.24s & (2) \\
$\delta$ (J2016) & +48$^{\circ}$ 22' 23.2'' & (2) \\
$\mu_\alpha \cos{\delta}$ & $-$1545.069$\pm$0.023\,mas\,a$^{-1}$ & (2) \\
$\mu_\delta$ & $-$962.724$\pm$0.054\,mas\,a$^{-1}$ & (2) \\
$V_{\rm r}$ & $-35\,609.2\pm9.8$\,m\,s$^{-1}$ & (3) \\
$d_*$ & 8.0426$\pm$0.0036\,pc & (2) \\
\noalign{\smallskip}
$G$ & 11.6847$\pm$0.0028\,mag & (2) \\
$J$ & 8.488$\pm$0.029\,mag &  (4) \\
\noalign{\smallskip}
$T_{\rm *,eff}$ & 3280$\pm$40\,K & (5)  \\
$\log{g}$ & 5.09$\pm$0.09\,dex & (5) \\
$[\sc{\rm Fe}/\sc{\rm H}]$ & $-0.12\pm0.10$ & (5) \\
$L_{\rm *,bol}$ & $(3315\pm18) \times 10^{-6}$\,L$_{\odot}$ & (6) \\ 
$R_*$ & 0.1781$\pm$0.0042\,R$_\odot$ & (7)  \\
$M_*$ & 0.1639$\pm$0.0093\,M$_\odot$ & (8) \\ 
\noalign{\smallskip}
$P_{\rm rot}$ & $140\pm10$\,d & (9) \\ 
$v \sin{i}$ & $<$2 km s$^{-1}$ & (10) \\
${\bf B}_*$ & $<680$\,G & (11) \\
$L_{\rm X}$ & $\sim 5.5 \times 10^{26}$\,erg\,s$^{-1}$ & (12)  \\
$\log{\it R'_{\rm HK}}$ & $-5.22\pm0.18$\,dex & (13) \\
pEW'(H$\alpha$) & $-0.08\pm0.15$\,\AA{} & (14)  \\
\noalign{\smallskip}
HZ & 0.047--0.124\,au & (15) \\
\noalign{\smallskip} \hline          
\end{tabular}
\tablefoot{
$^{(1)}$\cite{1996AJ....112.2799H},
$^{(2)}$\cite{2021A&A...649A...1G},
$^{(3)}$ this work, extracted from 95 CARMENES spectra with {\tt raccoon},
$^{(4)}$\cite{2003yCat.2246....0C},
$^{(5)}$\cite{2021A&A...656A.162M},
$^{(6)}$ this work, using {\it Gaia} DR3 data \citep{2021A&A...649A...1G} and following \cite{2020A&A...642A.115C},
$^{(7)}$this work, using Stefan-Boltzmann law, 
$^{(8)}$this work, using mass-radius relation by \cite{2019A&A...625A..68S},
$^{(9)}$this work, using MEarth and ASAS-SN photometry,
$^{(10)}$\cite{2018A&A...614A..76J},
$^{(11)}$\cite{2022A&A...662A..41R},
$^{(12)}$\cite{2020MNRAS.497.1015F},
$^{(13)}$this work, derived from the $R_{\mathrm{HK}}^\prime$ time series extracted from HARPS-N and ESPaDOnS spectra following \cite{2021A&A...652A.116P},
$^{(14)}$this work, following \citet{2019A&A...623A..44S},
$^{(15)}$recent Venus-early Mars habitable zones (HZ) following \cite{2013ApJ...765..131K}.\\
}
\end{table}

In this study, we analyse all the available spectroscopic and photometric data of GJ\,1151 in order to find evidence for the rotation period of the star, the mentioned short-period and long-period planets, and the source for the LOFAR radio signal. We carried out a detailed investigation of the influence of magnetic activity on the star, both in absolute values and in the variations over various timescales. We use further additional CARMENES and HARPS-N RVs as well as over 10 years of MEarth and ASAS-SN photometry. We present the observational data in Sect.\,\ref{sec:obsdata}. In Sect.\,\ref{sec:method} we explain the applied methodologies, and in Sect.\,\ref{sec:dataanal} we explain the analysis of the data. The results are shown in Sect.\,\ref{sec:res} and we conclude our study in Sect.\,\ref{sec:con}.

\section{Observational data} \label{sec:obsdata}

\subsection{Spectroscopic data}  \label{subsec:HPF}

The CARMENES instrument consists of a visible (VIS) and a NIR spectrograph with wavelength coverages of 0.52 to 0.96\,$\mu$m and 0.96 to 1.71\,$\mu$m, and resolutions $\mathcal{R}$ of 94\,600 and 80\,400, respectively \citep{2014SPIE.9147E..1FQ}. The VIS data used in this study consist of the 70 epochs already published in \citet{2021A&A...649L..12P} along with 31 additional observations. It is distributed over seven seasons, including 7 epochs from February 2016 to June 2016, 1 lone data point from January 2018, and 93 further epochs from February 2020 to April 2022, with a small gap from August to November. The basic reduction of the échelle spectra was done with {\tt caracal} \citep[CARMENES Reduction And CALibration software,][]{2016SPIE.9910E..0EC}. We further corrected the CARMENES VIS spectra for telluric lines using the method explained by \citet{2022nagel}. Radial velocities were then extracted with the {\tt serval} code considering nightly zero points and instrumental drifts (see Section\,\ref{subsec:spectro} for details). This dataset has the largest baseline of observations, relatively small uncertainties, and the lowest RV root mean square (rms). Two recent data points could not be considered in the calculation of the RVs because they showed S/N$<3$. One further data point is an obvious RV outlier and another shows uncertainties larger than 5\,$\sigma$. In the following, 97 CARMENES VIS epochs were considered. We list the measurements for each epoch in Table\,C.2.

\begin{figure*}
\centering
\includegraphics[width=\textwidth]{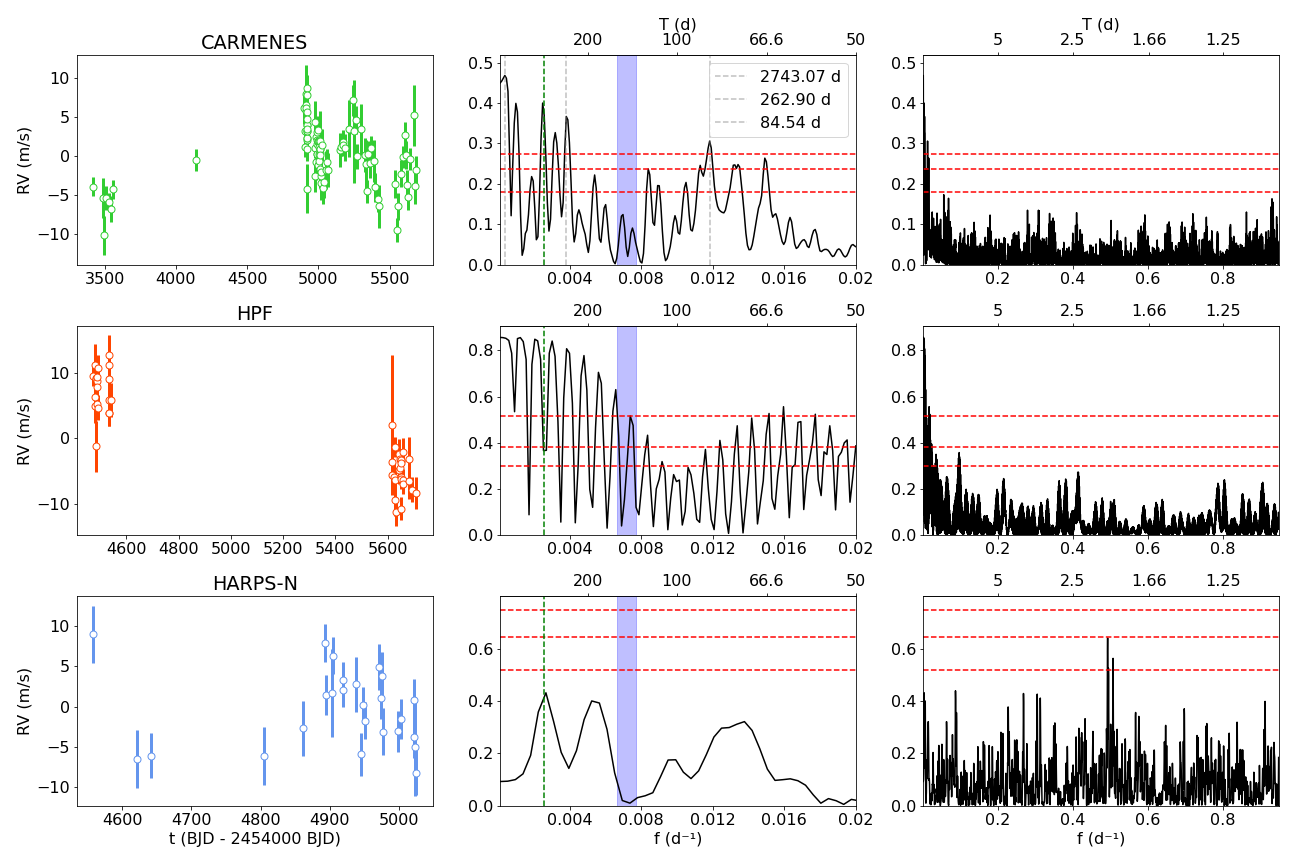}
 \caption{RV time series (left panels) and periodograms for the individual RV datasets coming from CARMENES, HPF, and HARPS-N (from top to bottom). We show the behaviour at longer periods $>$50\,days (middle panels) and down to 1\,day (right panels). We mark the rotation period of the star at 140$\pm$10\,d in blue, the orbital period of the proposed planet in green, and the prominent peaks in grey.} \label{individual_rv}
\end{figure*}

HARPS-N is a spectrograph observing at visible wavelengths, which is installed at the Nasmyth B focus of the 3.58\,m Telescopio Nazionale Galileo (TNG) at the Roque de Los Muchachos Observatory in La Palma, Spain \citep{2012SPIE.8446E..1VC}. HARPS-N has a wavelength coverage of between 0.38 and 0.69\,$\mu$m and a resolving power of $R\sim$115\,000. We use 21 publicly available data points already used in the previous studies along with 25 newly observed epochs from the A44TAC\_15 program. The dataset then consists of 46 epochs distributed over four seasons in two parts of approximately 3 months, from December 2018 to February 2019 and from February 2022 to April 2022. We extracted the RVs with both {\tt terra} and {\tt serval} pipelines and show the basic properties in Table\,\ref{tab:Rvdata} for completeness, but used the reduction by {\tt terra}, because it shows the lower rms and uncertainty. This comparison is not done for the CARMENES spectra because {\tt serval}  includes a variety of corrections especially designed for the spectra of the instrument. As we can observe in Fig.\,\ref{individual_rv}, the two different parts show a large offset of around 15\,m\,s$^{-1}$, resulting in the relatively large rms. We list the measurements for each epoch in Table\,C.1.

HPF is an NIR spectrograph installed at the 10\,m Hobby-Eberly Telescope at the McDonald Observatory in Texas, USA \citep{2012SPIE.8446E..1SM}. It has a wavelength coverage from 0.8 to 1.27\,$\mu$m and a resolution of $R\sim$55\,000. Mahadevan et al. (priv. comm.) kindly provided us with the data used in their previous study of the star \citep{2021ApJ...919L...9M}. These data consist of 25 epochs acquired from March 2019 to June 2020, computed from the reduced 1D spectra using {\tt serval}. We give an overview of the basic properties of the data in Table\,\ref{tab:Rvdata} and illustrate them in Fig.\,\ref{individual_rv}. As we can see, the HPF data have the shortest time baseline $T$, the shortest average sampling between observations $\Delta t$, and the largest average uncertainty $ \langle \sigma_{\rm RV} \rangle$. 

\begin{table*}
\caption{Basic characteristics of the RV data for the different instruments and the combined dataset.} \label{RVsum}  
\centering  
\begin{tabular}{lccccc}
\hline \hline \noalign{\smallskip} 
        & HPF & \multicolumn{2}{c}{HARPS-N} & CARM VIS & combined       \\ 
        &     & {\tt terra} & {\tt serval} & \\
\noalign{\smallskip} \hline \noalign{\smallskip} 
$N$ & 25 & \multicolumn{2}{c}{46} & 97 & 168 \\
rms [m\,s$^{-1}$] & 4.64 & 6.34 & 7.36 & 3.81 & 4.82 \\
$ \langle \sigma_{\rm RV} \rangle$ [m\,s$^{-1}$] & 3.00 & 1.89 & 2.71 & 1.94 & 2.82 \\
$T$ [d] & 468 & \multicolumn{2}{c}{1236} & 2268 & 2290\\
$\Delta t$ [d] & 19.5 & \multicolumn{2}{c}{27.5} & 23.6 & 7.27 \\
date & Mar19--Jun20 & \multicolumn{2}{c}{Dec18--Apr22} & Feb16--Apr22 & Feb16-Apr22 \\
\noalign{\smallskip} \hline
\end{tabular}
\tablefoot{Number of observations $N$, RV root mean square (rms) of the observed data, average internal RV uncertainties $\langle \sigma_{\rm RV} \rangle$, time baseline $T;$ and average time interval between observations $\Delta t$ for the RV data for the different instruments and the combined dataset.} 
\label{tab:Rvdata}
\end{table*}

We further analyse 60 spectropolarimetric observations from ESPaDOnS\footnote{Publicly available at \url{ http://polarbase.irap.omp.eu}.} \citep[Echelle SpectroPolarimetric Device for the Observation of Stars;][]{2003ASPC..307...41D}. This is a high-resolution échelle spectrograph operating in the wavelength range from 0.37 to 1.05\,$\mu$m with a resolution of $\mathcal{R}=68\,000$, which is mounted at the CFHT (Canada-France-Hawaii-Telescope). GJ\,1151 was observed over the course of six different nights and on ten different occasions from January to March 2017 (four Stokes I observations for one Stokes V measurement -- the polarisation filter is in different angles). The dates are contemporaneous with the RV monitoring observations. We use the ESPaDOnS spectra to access the longitudinal stellar magnetic field and to extract $\log R'_{\rm HK}$ values (see Table\,\ref{table:GJ1151data}) following the procedure in \citet{2021A&A...652A.116P}. The RVs extracted from the spectra are not used because they show large uncertainties of approximately 15\,m\,s$^{-1}$.

\subsection{Data extraction from spectroscopic observations}  \label{subsec:spectro}

There are two main methods to extract RVs from high-resolution échelle spectra. The binary-mask technique was developed early on and implemented in the data-reduction system (DRS) of both HARPS and HARPS-N. This method computes the Doppler shift via the cross-correlation function (CCF) between the stellar spectrum and that of a predefined binary mask. The resulting shape of the CCF is measured through its moments, namely its width and height (FWHM -- full width half maximum, and CON -- contrast), and its asymmetry, the bisector inverse span  (BIS). Those indices have been shown to be influenced by the effects of stellar activity on the spectra. The CCF technique \citep{1974A&A....31..129S, 2021MNRAS.507.1847R, 2022MNRAS.512.5067K} is also implemented by the {\tt raccoon} \cite[Radial velocities and Activity indicators from Cross-COrrelatiON with masks,][]{2020A&A...636A..36L} code, which we also use here. The {\tt raccoon} pipeline for GJ\,1151 uses a mask of spectral type (M4.5\,V) derived from the CARMENES observations.

A second method, dubbed template matching, has proven to yield better results than the CCFs \citep{2012ApJS..200...15A}, especially for low-mass stars, where many of the abundant spectral lines are blended. In such cases, a template spectrum of high signal-to-noise ratio (S/N) is constructed by combining all the observed spectra of a target. Each observed spectrum is then optimally matched to this template. This method constitutes the basis of the {\tt terra} \citep[Template-Enhanced Radial velocity Re-analysis Application,][]{2012ApJS..200...15A} and {\tt serval} \citep[SpEctrum Radial Velocity AnaLyser,][]{2018A&A...609A..12Z} codes. 

The {\tt serval} code includes the calculation of two activity indicators, the differential line width (dLW), measuring the width of the average line, similar to CCF FWHM, and the chromatic index (CRX), which is the slope over the logarithmic wavelength of the RVs calculated for every order of the échelle spectrum \citep{2020A&A...641A..69B}. Further, the CARMENES processing pipeline delivers measurements for a variety of line features sensitive to stellar activity by comparing their fluxes with the continuum values. Those include the H$\alpha$ line at 6\,563\,\AA{}, which can trace the ionised hydrogen content, and the Ca\,{\sc ii} infrared triplet (CaIRT). For this index, the three ionised calcium emission lines of wavelengths 8\,498, 8\,542, and 8\,662\,\AA{} are measured. We then create the CaIRT index as the average over the three individual measurements. We further calculate the sodium line indices Na\,{\sc i}\,D2 and Na\,{\sc i}\,D1 at vacuum wavelengths of 5\,890 and 5\,896\,\AA{}, and use the average to create  the Na\,{\sc i}\,D index for the line doublet.

We list the CCF activity indices calculated by the {\tt raccoon} code, the described indices calculated by {\tt serval}, and the H$\alpha$ index in Table\,C.2 for the CARMENES spectra and in Table\,C.1 for HARPS-N. Additionally, we show the calcium and sodium indices calculated from the CARMENES spectra. Five HARPS-N data points including publicly available spectra showed problems in the calculation of the CCF due to low S/N.

\subsection{Photometric data}  \label{subsec:MEarth}

MEarth-North is an array of eight 40cm telescopes located at the Fred Lawrence Whipple Observatory (FLWO) on Mount Hopkins, Arizona, USA \citep{2008AAS...212.4402C}, performing photometric monitoring of the closest late M dwarfs since 2008. Our target has been observed with \emph{telescope 2,} producing two different datasets from 2008 to 2010 and from 2011 to 2020, and with \emph{telescope 7} from 2011 to 2020. We provide an overview of the basic properties of the datasets in Table\,\ref{Photsum}. We also show the combined data in Fig.\,\ref{LS_indices} and the individual data in the Appendix in Fig.\,\ref{prewhit1}, after the application of a daily binning and a 3$\sigma$ clipping of the data.

ASAS-SN (All-Sky Automated Survey for SuperNovae) is a survey instrument composed of five stations, two of which are located at Cerro Tololo International Observatory (CTIO, Chile), and one each at Haleakala Observatory (Hawaii), McDonald Observatory (Texas), and South African Astrophysical Observatory (SAAO, Sutherland, South Africa) \citep{2017PASP..129j4502K}. As its automated data pipeline is not optimised for the observation of high-proper-motion stars, we calculated new coordinates prior to the download of the light curves for each year of observations. We calculated those coordinates at the middle of every season of observations (April) following the procedure by \citet{2021Sci...371.1038T}. We analysed observations in the $V$ band from July 2013 to January 2019 and in the $g$ band from September 2017 to June 2022. The latter set includes a linear trend $(43.5\pm4.8) \times 10^{-6}$\,mag\,d$^{-1}$, which we subtracted in order to improve the sampling of the period range we are interested in. We show the nightly binned and 3$\sigma$-clipped data in Fig.\,\ref{prewhit1} and their basic properties in Table\,\ref{Photsum}.

We also inspected the TESS mission photometry of GJ\,1151, which was observed from 23 February to 17 March 2020 (Sector 22) and from 30 January to 24 February 2022 (Sector 48). As shown already by \citet{2021ApJ...919L..10P}, no rotation modulations (the rotation period is not covered by the 27day TESS sectors), planetary transits, or obvious flares can be found. 

Furthermore, we collected photometry at 37 epochs (August 2019 to May 2022) from 320 observations from the {\it Gaia} DR3 database \citep{2021A&A...649A...1G} after performing a retrieval on its `Variable dataset of Solar like stars' (stars with spectral type later than F5). We could not identify any significant signals in the noisy periodogram of the low-cadence {\it Gaia} data.

\begin{table*}
\caption{Basic characteristics of the nightly binned and 3$\sigma$-clipped photometric data.} \label{Photsum}  
\centering  
\begin{tabular}{lccccccc}
\hline \hline \noalign{\smallskip} 
        &  \multicolumn{4}{c}{MEarth} & \multicolumn{2}{c}{ASAS}        \\ 
        &  Tel2 A  & Tel2 B & Tel7 & combined & $g$ band & $V$ band & \\
\noalign{\smallskip} \hline \noalign{\smallskip} 
$N$ & 82 & 320 & 241 & 638 & 293 & 280 &  \\
rms [mmag] & 3.72 & 7.07 & 5.71 & 6.51 & 49.04 & 24.27 &  \\
$ \langle \sigma_{\rm phot} \rangle$ [mmag] & 0.89 & 1.46 & 1.01 & 1.21 &  18.07 & 13.28 & \\
$T$ [d] & 227 & 3045 & 2686 & 4137 & 1656 & 2124 & \\
$\Delta t$ [d] & 2.8 & 9.5 & 11.2 & 6.5 & 5.7 & 7.6 \\
date & Nov08--Jun09 & Nov11--Mar20 & Nov12--Mar20 & Nov08--Mar20 & Nov17--May22 & Feb13--Nov18\\
\noalign{\smallskip} \hline 
\end{tabular}
\tablefoot{MEarth combined data are shown under the "combined" column. $ \langle \sigma_{\rm phot} \rangle$ refers to the photometric uncertainty. See notes in Table\,\ref{RVsum} for more details.} 
\label{table instruments}
\end{table*}

\section{Methodology}  \label{sec:method}

\subsection{Periodogram analysis}  \label{subsec:freq}

To search for periodic signals in our data, we used the Generalised Lomb Scargle (GLS) periodogram as described in \citet{2009A&A...496..577Z}. The significance of the signals in the periodograms is computed with the false-alarm probability (FAP) using the bootstrap method \citep{efron1985bootstrap}. We consider the signals to be tentative or significant if their FAP is below the 1$\%$ or the 0.1$\%$ level, respectively. We perform a pre-whitening process, which consists of fitting the data with a sine curve with the same period and phase as the most significant peak in the periodogram. This signal is removed and a re-analysis is performed on the residuals until no further significant signals are found in the data.

\subsection{Model comparison}  \label{subsec:stats}

As an alternative approach for analysing the data, we searched for a best-fitting model through Bayesian statistics \citep{2008ConPh..49...71T}, as implemented in {\tt juliet} \citep{2019MNRAS.490.2262E}. Starting with the simple combination of datasets (null model, e.g. for the MEarth dataset combination), we calculate different models of increasing complexity and accept a more complex model over a simpler one as significant if the difference of the logarithm of the Bayesian evidence (or log-evidence) between the more complex and simpler model is $\Delta \ln{\rm Z} > 5$, as discussed in \citet{2008ConPh..49...71T}. The Bayesian evidence is the integral of the maximum likelihood estimator over the prior volume, which is the parameter space over which we evaluate our models and that we choose to be as large as possible. We calculate $\Delta \ln{\rm Z}$ with {\tt juliet} using the dynamic nested sampler {\tt dynesty}, which iteratively evaluates the integral of the maximum likelihood over different live points. Those with poorest likelihood are replaced by randomly chosen points that improve the Bayesian evidence, until the improvement is below a certain threshold. As a dynamic nested sampler, the number of live points that are used is also updated on the fly, with a starting number of 1\,000. This methodology allows us to achieve precise posterior distributions of every parameter of the model within the limits of the introduced priors thanks to the efficient exploration of the prior space.

\subsection{Model definitions}  \label{subsec:mod}

We fit a variety of different models, $M$, to our data. In the following, we describe each of the components from which the models are constructed including their respective parameters and hyper-parameters, which we generally evaluate from wide uninformative priors.

\begin{itemize}

    \item Keplerian and sinusoid curve:
    \begin{equation}
     M_{\rm sk,i} = K_i \cdot [\cos ({\mathbf{\omega_i}+\nu_i(\tau)}) + e_i \cos{\mathbf{\omega_i}}],
    \end{equation}    
    with $\tau=t-t_{0,i}$, which is the observed time minus an initial time $t_{0,i}$ of model $i$, with $K_i$ being its amplitude, $P_i$  the period, $\omega_i$  the argument of periastron, $e_i$  the eccentricity, and $\nu_i$  the true anomaly of model $i$. The periods are evaluated from 1\,d up to two times the baseline of our observations, $2T$. We apply a uniform prior on the log-scale of the observed times so that we can equally explore the different timescales of our problem. The amplitude is uniformly explored up to $2 {\rm rms}_{\rm RV}$, the eccentricity is explored with a uniform prior from 0 to 1, and $\omega$ from 0 to 360\,deg. A sinusoid is modelled in the case of $e_i=0$. Other tests with wider priors on $P_i$ and $K_i$ were explored in order to avoid missing very eccentric signals. These solutions at periods much larger than $T$ were highly dependant on the offsets between datasets.
    
    \item Individual offsets, linear and quadratic terms:
    \begin{equation}
     M_{\rm pol,j} = \gamma_{j} + \dot{\gamma} \tau + \ddot{\gamma} \tau^2,
    \end{equation}
    where $\gamma_{\rm j}$ is the offset of the $j$-th dataset (e.g. for different instruments CV, HA, HP for CARMENES, HARPS, HPF), and $\overset{.}{\gamma}$ and $\overset{..}{\gamma}$ are the linear and quadratic coefficients of a parabola. We apply uniform priors on the offsets from $-2 {\rm rms}_{\rm RV}$ to $2 {\rm rms}_{\rm RV}$. For the linear and quadratic term, we set uniform priors from $-2 {\rm rms}_{\rm RV} T^{-1}$ to $2  {\rm rms}_{\rm RV} T^{-1}$ and $-2 {\rm rms}_{\rm RV} T^{-2}$ to $2 {\rm rms}_{\rm RV} T^{-2}$, respectively, where $T$ is the time baseline of the observations.
    
    \item Individual jitter: \\ 
    Every dataset is assigned an individual jitter term $\sigma_{j}$ representing the relative significance between different datasets $j$ and accounting for possible underestimations of the individual uncertainties. The jitter is added quadratically to the individual errors $\sigma_{\rm RV}$ of each epoch in the calculation of the model likelihoods. Similarly to the offsets, we apply uniform priors of up to $2 {\rm rms}_{\rm RV}$.
    
    \item Gaussian processes:\\
    Our modelling of the stellar contribution to the datasets considers the usage of a specific kernel $S$ in the application of a Gaussian process regression. Those kernels are able to reproduce the effect of inhomogeneities on the stellar surface, and have a quasi-periodic form, where the signal of the rotation is allowed to change its parameters, namely amplitude and phase, over a certain time interval. We applied the quasi-periodic plus cosine (QPC) kernel, as now included in {\tt juliet}, with the following form \citep{2021A&A...645A..58P}:
    \begin{eqnarray}
    S(\Delta\tau) = {\rm exp}\Big(-2 \frac{\Delta\tau^2}{\lambda^2}\Big) \cdot \nonumber \\
    \Big[ h_1^2 {\rm exp} \Big( - \frac{1}{2 \omega_0^2} \sin^2 (\pi \frac{\Delta\tau}{P}) \Big)
    + h_2^2 \cos(4 \pi \frac{\Delta\tau}{P}) \Big],
    \end{eqnarray}
    where $\lambda$ is the exponential decay representing the average lifetime of starspots, $K=\sqrt{h_1^2+h_2^2}$ is the average amplitude of the signal, $w_0=0.31$ is a constant, and $P$ represents the rotational period of the star. We set a log-uniform prior on $\lambda$ from 60\,d to the time baseline $T$ of our observations. For $P$, we use wide uninformative priors.     

\end{itemize}

\section{Data analysis} \label{sec:dataanal}

\subsection{Photometry} \label{sec:photometry}

\begin{figure*}
\centering
\includegraphics[width=\textwidth]{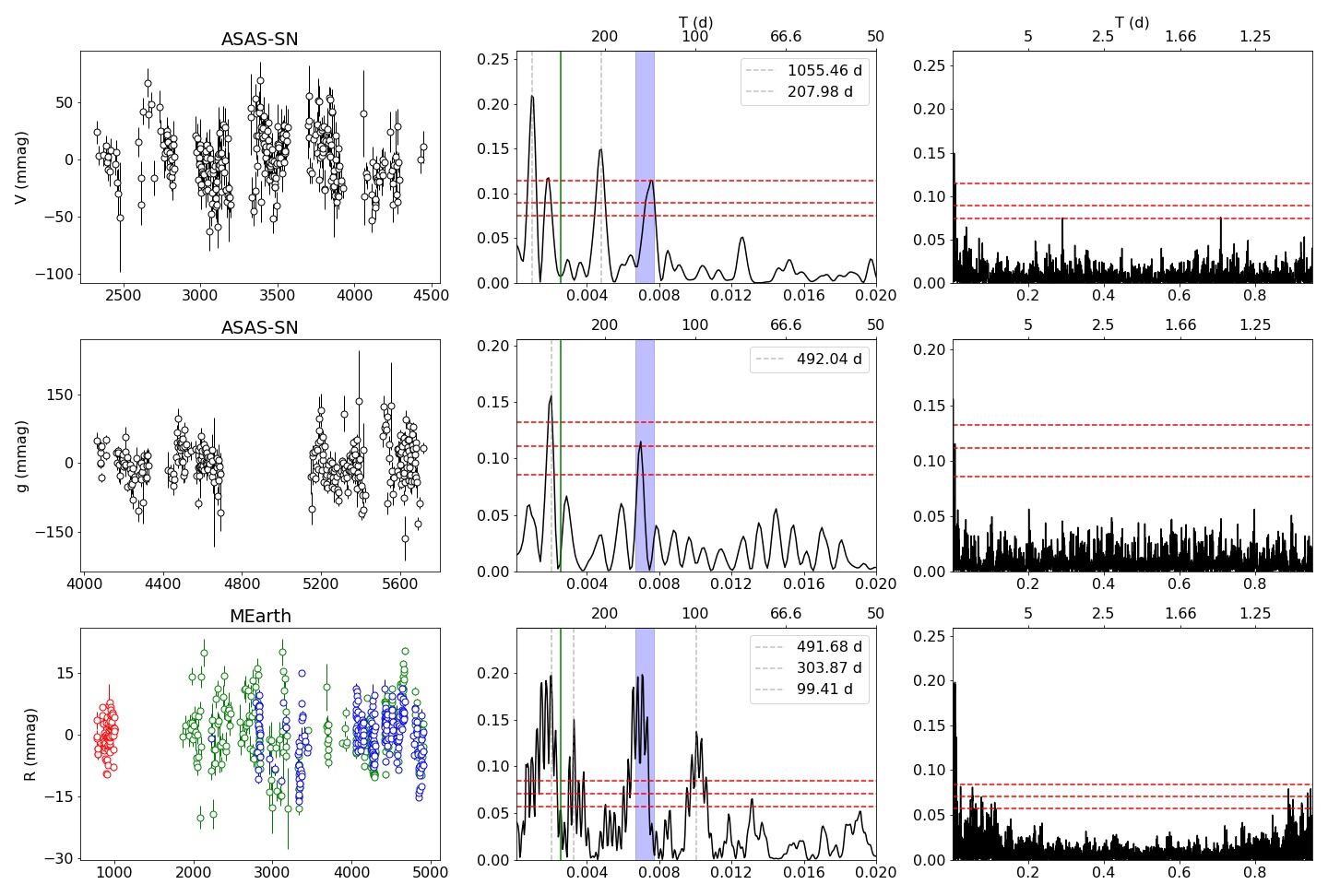}
\includegraphics[width=\textwidth]{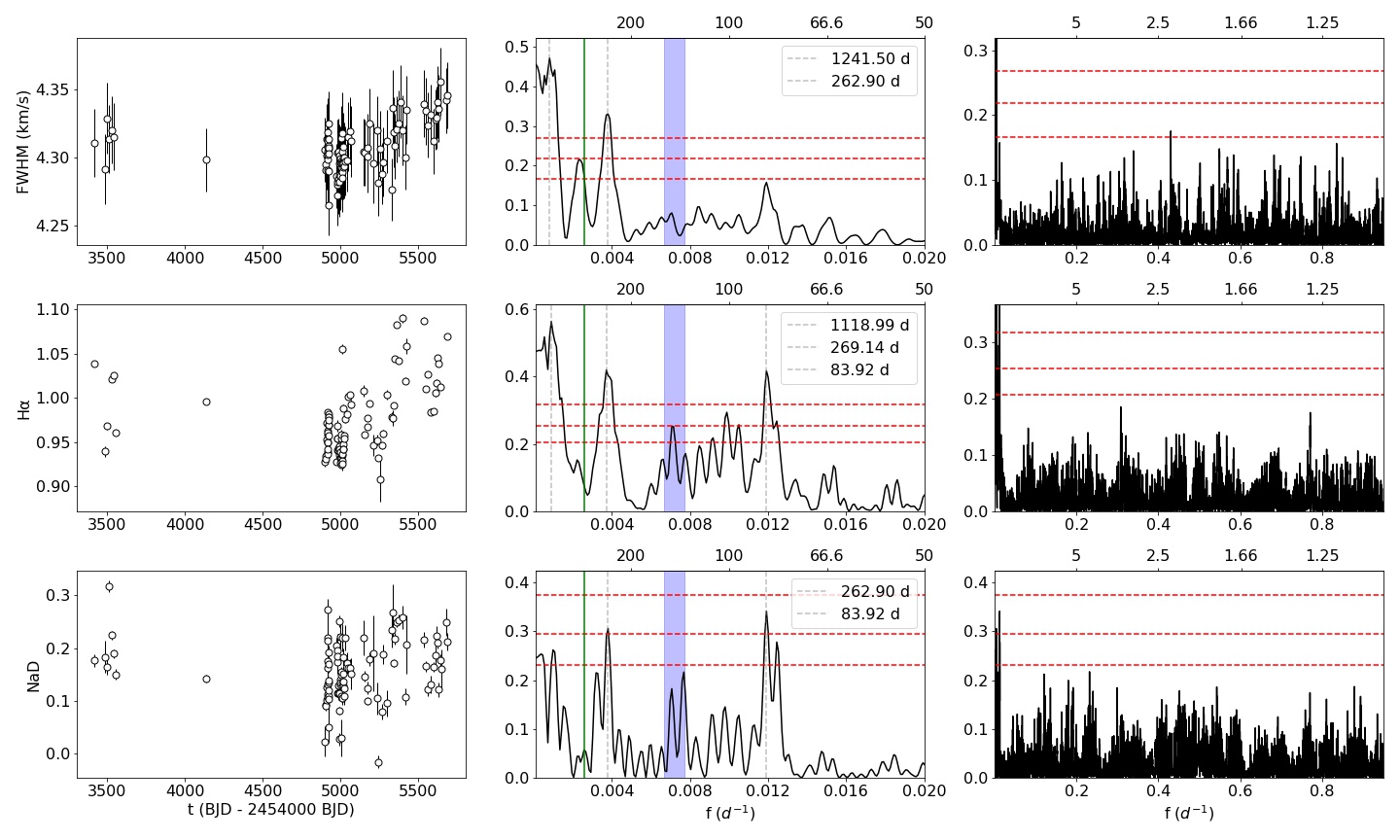}
\caption{Photometric and spectral index time series. Significant signals are marked. We show the full dataset (left panels) and their periodograms for longer periods ($>$50\,d, middle panels) and for all periods (right panels). The grey vertical dashed lines mark the significant peaks. We show the period of the planet  in green and the rotational period of the star at 140$\pm$10\,d  in blue. For the combined MEarth data, we show the different individual sets in colours. The red dashed horizontal lines show the FAP levels of 10, 1, and 0.1\%.} \label{LS_indices}
\end{figure*}

We show in Figs.\,\ref{LS_indices} and \ref{prewhit1} the data and periodograms for different individual and combined photometric datasets. In the ASAS-SN data, we identify long-term modulations at 490 and 1055\,d, and prominent signals around 140\,d and 210\,d, which may be connected by yearly aliasing, that is $(140^{-1}-365^{-1})^{-1}$. The MEarth dataset from 2008 to 2010 is the basis for the previously published rotational periods of GJ\,1151 of 125\,d \citep{2019A&A...621A.126D} and 117.6\,d \citep{2016ApJ...821...93N}. It can be seen in the additional MEarth sets that the signal actually appears at a significantly longer period ($\sim$140\,d), and that yearly aliases of this period are visible at around 220 and 100\,d. Therefore, we update the rotational period to 140$\pm$10 d, setting a conservative error bar that includes the prominent signals observed in the periodograms of the ASAS-SN and MEarth data. We performed a combined analysis of the three different MEarth datasets where we searched for the offsets and jitters for each set that maximise the log-evidence (see Sect.\,\ref{subsec:mod}). Besides the main signal at around 140\,d, we also identify a modulation of 500\,d. We attempted to fit a GP model with a QPC kernel to the photometry, but the process is not able to converge on a specific period or lifetime of the spots. Most importantly, we do not see any periodicity at $\sim390$\,d in the photometric data, which is the period for the RV signal of the potential planetary companion.

\subsection{Stellar activity indices} \label{sec:activity}

Figure \ref{LS_indices} shows the spectroscopic indices that have significant signals calculated from CARMENES and HARPS-N spectra using {\tt serval}, {\tt raccoon}, and {\tt terra} as discussed. All other indices are shown in the Appendix in Figs.\,\ref{prewhit2} and \ref{prewhit3}. The distribution of peaks in the periodograms of the FWHM, H$\alpha,$ and Na\,{\sc i}\,D indices from CARMENES observations is similar to that found in the photometry. All of these indices show long-term variability at $>1000$\,d, and possibly the two yearly aliases of the 140d rotation period at around 230 and 100\,d. On the other hand, no index shows a clear signal at 390\,d or at 140\,d.

\subsection{Radial velocities} \label{sec:RV}

The RVs of the different datasets are shown in Fig.\,\ref{individual_rv}. We can see that the periodogram of the CARMENES data contains multiple peaks with large power at $>200$\,d, including a long-term modulation of the order of the time baseline of the observations ($>2000$\,d). The periodogram of the HPF data does not show any significant signal except for the tentative variability at around 2\,d reported by \citet{2021ApJ...919L...9M}, which was later found to be caused by the poorly sampled long-period signal \citep{2021A&A...649L..12P}. The periodograms of the sparsely sampled HARPS-N datasets are dominated by the window function and do not show clear periodogram peaks. No RV dataset shows significant power at the rotation period of 140\,d.

We fit different models with increasing level of complexity to the three RV datasets in order to optimally combine the data and to find significant signals in the full data set. Each of the parameters of the models is evaluated over wide uninformative priors. All the models should represent either a stellar or a planetary signal imprinted on the measurements. We show the log-evidence values of all models as evaluated by the {\tt juliet} code in Table\,\ref{table_lnZ}.

\begin{table}
\caption{Bayesian evidences of the different models calculated for the three combined RV datasets from CARMENES, HARPS-N, and HPF instruments.}
\centering  
\begin{tabular}{l c c}
\hline \hline \noalign{\smallskip} 
model   &        $\ln{\rm Z}$ & $\Delta \ln{\rm Z}$     \\
\noalign{\smallskip} \hline \noalign{\smallskip}
null & $-506.6\pm$0.4 & $69.1\pm$1.1 \\
sinusoid & $-483.1\pm$0.7 &  $45.6\pm$1.2 \\
Keplerian & $-458.3\pm$0.7 & $20.8\pm$1.2 \\
{\bf sinusoid + parabola} & {\bf $-$441.7$\pm$0.8} & {\bf 4.2$\pm$1.3} \\
Keplerian + parabola & $-443.1\pm$0.8 & $5.6\pm$1.3 \\
sinusoid + sinusoid & $-440.0\pm$0.8 & $2.5\pm$1.3 \\
sinusoid + Keplerian & $-437.5\pm$1.0 & $0$ \\
Keplerian + sinusoid & $-441.2\pm$0.8 & $3.7\pm$1.3 \\
Keplerian + Keplerian & $-440.0\pm$1.0 & $2.5\pm$1.4 \\
\noalign{\smallskip} \hline
\end{tabular}
\tablefoot{$\Delta \ln{\rm Z}$ refers to the difference in log-evidence of the best model compared to any other model. The first word in each line of the first column refers to the variation with the shortest period (around 390\,d), the second word to the variation with the largest period. The model we chose to be most relevant for this study is shown in bold face.} 
\label{table_lnZ}
\end{table}

\begin{figure*}
\centering
\includegraphics[width=\textwidth]{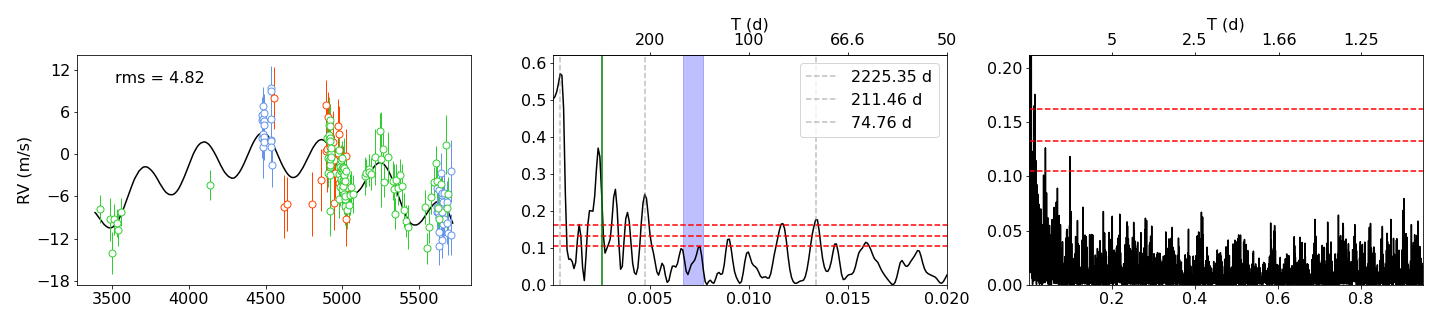}
\includegraphics[width=\textwidth]{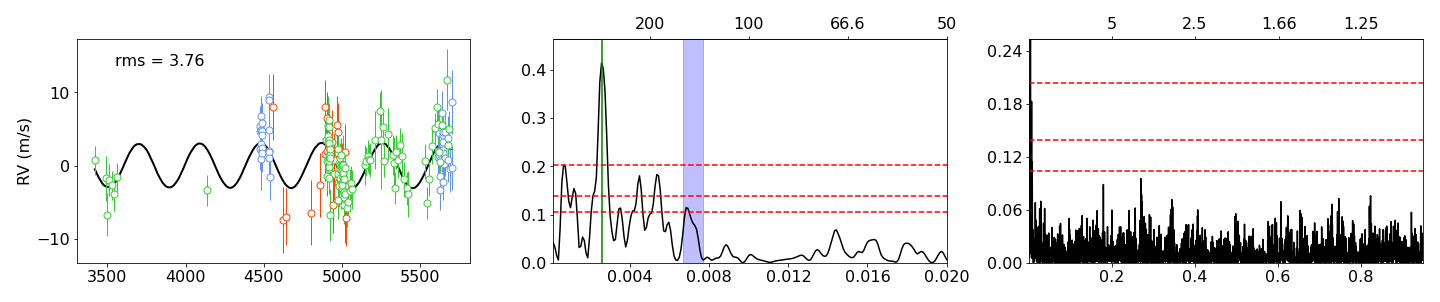}
\includegraphics[width=\textwidth]{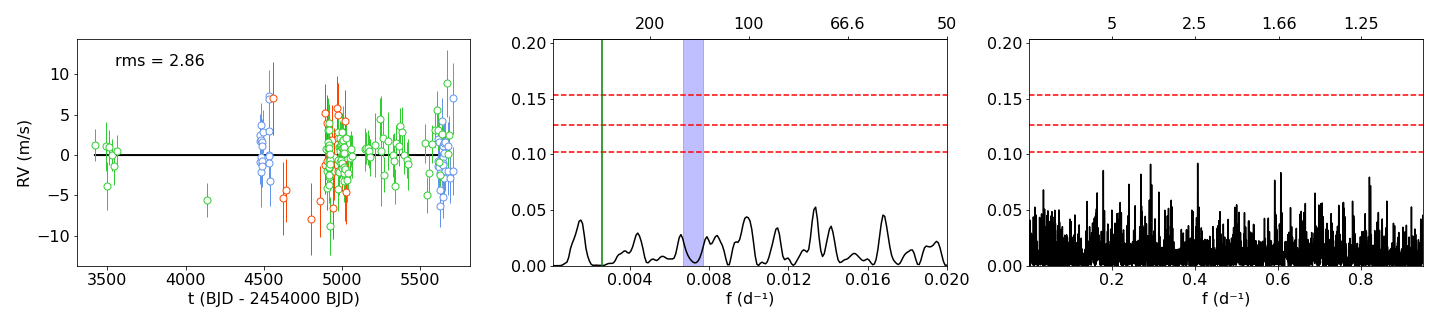}
 \caption{RV time-series data (left panels) and periodograms (red dashed lines as the 0.1, 1, and 10\% FAP levels) of the best-fitting model consisting of a sinusoid and a parabola. The middle panels show the low-frequency portion of the periodograms ($P>50$\,d) and we mark the rotational period at 140,\, including its uncertainty of $\pm$10\,d, with the blue vertical area, and the planetary period at 390\,d with the green vertical line. In the right panel, the full periodogram is shown. In the top panels, we show the data as observed (green: CARMENES, red: HPF, blue: HARPS-N) and indicate the best-fitting model by the black curve. The middle row shows the RV residuals after the subtraction of the parabola. The bottom row shows the residuals of the best fit and we provide the RV rms.} \label{fig:best_models1}
\end{figure*}

We obtain a value of $\ln{\rm Z}= -437.5$ as the largest log-evidence. Models with $\Delta \ln{\rm Z} <5$ with respect to this model, $\ln{\rm Z} < -442.5$, are statistically equivalent. All such models have both a common, strong signal at $\sim$390\,days and a second long-term signal. Given our statistical threshold, the 390d signal is best fitted by a sinusoid. The second signal cannot be as clearly identified because the periods found from both sinusoid and Keplerian models are below the Nyquist frequency of $f=0.0087$\,d$^{-1}$ ($P=0.5 T \approx 1145$\,d) and do not converge on a specific period. This is also true for the hyper-parameters of the fitted QPC kernel in the GP approach, which also do not converge and which we therefore do not show in the table. As a result, we prefer the less complex model and fit the second signal using a parabola. We show the combined time series and the corresponding periodograms in Fig.\,\ref{fig:best_models1}.

\section{Results} \label{sec:res}

\subsection{Detection of a sub-Neptune planet}  \label{subsec:GJ1151b}

\begin{figure}
  \centering
  \includegraphics[width=9cm]{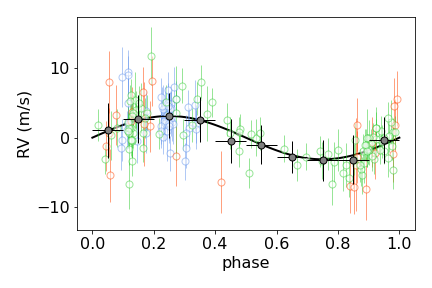}
  \caption{Phase-folded RV data of the sub-Neptune planet orbit for the sinusoid + parabola model with the parabola term subtracted. Symbols are coloured as in Fig.\,\ref{fig:best_models1}. Black dots show the RV average of each 0.1\,phase bin.} \label{fig:best_models_phase}
\end{figure}

Following the criteria of an improvement in $\ln{\rm Z}>5$, the best-performing model is the combination of a sinusoid with a parabola (Table \ref{table_lnZ}). Due to the lack of similar periodicities in any of our stellar activity tracers (Section \ref{sec:activity}), we conclude that the sinusoidal modulation at 390\,days is most likely produced by a planetary companion in a nearly circular orbit. The strength of the signal, its clear visibility in the RV curve, and our thorough RV extraction make it furthermore very unlikely for the signal to be introduced by any yearly effects \citep[as reported for HARPS data in][]{2015ApJ...808..171D}. We show all fitted parameters of this model in Table\,\ref{tab:params}, including the detailed priors and additionally derived parameters. A corner plot of the posterior distribution for those parameters is shown in the Appendix in Fig.\,\ref{corner_sin_quad}. Following those parameters, GJ\,1151\,b appears to be a sub-Neptune mass planet with a minimum mass of 10.6\,M$_\oplus$, and in an orbit of non-measurable eccentricity. The orbital distance is 0.57\,au, which places the planet well outside the habitable zone around GJ\,1151. We further show the RV residuals after the application of the preferred model and the corresponding periodograms in Fig.\,\ref{fig:best_models1}. The RV measurements  are shown in Fig.\,\ref{fig:best_models_phase} after subtracting the parabola and being phase-folded to the planetary orbital period.

\begin{table*}
\caption{Parameters for the best-fit model of a circular orbit and a second signal represented by a second-order polynomial.}
\centering  
\begin{tabular}{lccl}
\hline \hline 
\noalign{\smallskip} 
Parameter       & Units & Sinusoid + parabola & Priors  \\
 & & Model & \\
\noalign{\smallskip} 
\hline 
\noalign{\smallskip}
\multicolumn{4}{c}{\it Fitted parameters} \\
\noalign{\smallskip}
$P_1$ & d & 389.7$^{+5.4}_{-6.5}$ & $\log \mathcal{U} (1,2 T)$\\
$K_1$ & m\,s$^{-1}$ & 3.10$^{+0.38}_{-0.43}$ & $\mathcal{U} (0,s)$\\
$t_{0,1}$ & d & $t_{\rm 1}$+379$^{+22}_{-20}$ & $\mathcal{U} (t_{\rm 1}+200, t_{\rm f}+600)$\\
\noalign{\smallskip} 
$\dot{\gamma}$  & $10^{-4}$\,m\,s$^{-1}$\,d$^{-1}$ & 5.8$^{+5.2}_{-5.0}$ & $\mathcal{U}(-2  s T^{-1}, 2 s T^{-1})$ \\
$\ddot{\gamma}$ & $10^{-6}$\,m\,s$^{-1}$\,d$^{-2}$ & $-$7.54$^{+0.59}_{-0.61}$ & $\mathcal{U}(-4  s T^{-2}, 4 s T^{-2})$ \\
\noalign{\smallskip} 
$\gamma_{\rm CV}$ & m\,s$^{-1}$ & 3.85$^{+0.47}_{-0.52}$ & $\mathcal{U}(-s,s)$\\
$\gamma_{\rm HA}$ & m\,s$^{-1}$ & 4.75$^{+0.71}_{-0.71}$ & $\mathcal{U}(-s,s)$\\
$\gamma_{\rm HP}$ & m\,s$^{-1}$ & 1.00$^{+0.90}_{-0.88}$ & $\mathcal{U}(-s,s)$\\
$\sigma_{\rm CV}$ & m\,s$^{-1}$ & 1.55$^{+0.27}_{-0.28}$ & $\mathcal{U}(0,s)$\\
$\sigma_{\rm HA}$ & m\,s$^{-1}$  & 2.07$^{+0.47}_{-0.43}$ & $\mathcal{U}(0,s)$\\
$\sigma_{\rm HP}$ & m\,s$^{-1}$ & 2.71$^{+0.97}_{-1.02}$ & $\mathcal{U}(0,s)$\\
\noalign{\smallskip} 
\multicolumn{4}{c}{\it Calculated values} \\
\noalign{\smallskip} 
$M_1 \sin{i}$ & M$_\oplus$ & 10.62$^{+1.31}_{-1.47}$ & ... \\
a$_1$ & au & 0.5714$^{+0.0053}_{-0.0064}$ & ... \\
\noalign{\smallskip} 
\hline
\end{tabular}
\tablefoot{$s = 3 \times {\rm rms}_{\rm RV}$; $T$ is the time baseline of 2268\,days; $t_{\rm 1}$ the first epoch of observations of BJD = 2457419.67\,d; HA, HP, CV refer to the instruments HARPS-N, HPF, and CARMENES VIS. The parameters are shown as introduced in Sect.\,\ref{subsec:mod}.}
\label{tab:params}
\end{table*}

\subsection{Activity level of GJ\,1151}  \label{sec:halpha}

GJ\,1151 is a slow rotator with an inferred rotation period of $P=140\pm10$\,d (Section \ref{sec:photometry}) and projected equatorial velocity $v\sin i< 2$\,km\,s$^{-1}$. Its X-ray luminosity is $L_{\rm X} \sim  5.5 \times 10^{26}$\,erg\,s$^{-1}$ (see Table\,\ref{table:GJ1151data} for properties and related references). We calculate a low value of $\log R'_{\rm HK}=-5.22\pm0.18$\,dex using the ESPaDOnS observations from early 2017. We find published pEW(H$\alpha$) values of $0.0\pm0.2\,\AA{}$ from the CAFE spectrograph \citep[Calar Alto Fiber-fed Echelle;][]{2018A&A...614A..76J} and $-0.095\pm0.013\,\AA{}$ as an average of CARMENES measurements \citep{2019A&A...623A..44S}. With the additional epochs of observations acquired with CARMENES, we update this value to an average of $-0.08 \pm 0.15\,\AA{}$. GJ\,1151 can be classified as inactive when adopting a threshold of $-0.3$ \AA{} \citep{2021A&A...652A..28L} to separate magnetically active from inactive stars; but H$\alpha$ emission was detected by \citet{2018A&A...612A..49R}, with $\log{L_{\rm H \alpha}}/ L_{\rm bol} = -4.75$. \cite{2022A&A...662A..41R} reports an upper limit to the average surface magnetic field of GJ\,1151 of ${\bf B} < 680$\,G, comparing (for hundreds of M dwarfs) the observed broadening of different magnetically sensitive lines to the polarised radiative transfer calculations. With the Stokes V (circularly polarised) detection diagnosis \citep{1997MNRAS.291..658D}, the spectropolarimetric ESPaDOnS observations were able to marginally detect its magnetic field, with one epoch being on the verge of a definite detection (see Table\,\ref{tab:espadons}). We compute the values for the longitudinal magnetic field, ${\bf B_l}$, using the method described in \citet{2022MNRAS.514.4300B}.

On the other hand, the residual RVs (after the subtraction of our best model) still present a scatter of $\sim 3$\,m\,s$^{-1}$. This is rather high given the median uncertainty of the CARMENES data, $1.8$\,m\,s$^{-1}$, and the reported experiences when using the CARMENES survey to investigate these kinds of targets \citep{2022ribas}. From mid-2021 (BJD$\sim 2459400$) onwards, the star also seemed to enter into a more active phase as can be seen in the H$\alpha$ indices as depicted in Fig.\,\ref{LS_indices} for CARMENES and in Fig.\,\ref{prewhit3} for HARPS-N measurements, and by the increase of the RV residual scatter in the latest observations (see Fig.\,\ref{fig:best_models1}). We also find signals of long-term variability in some activity indices and photometry, which could be induced by such an increase in magnetic activity.

\begin{table}
\caption{Longitudinal magnetic field extracted from 60 ESPaDOnS spectropolarimetric observations.}
\centering  
\begin{tabular}{ccc}
\hline \hline 
\noalign{\smallskip} 
BJD [d] & ${\bf B_l}$ [G] & detection \\
\noalign{\smallskip} \hline \noalign{\smallskip} 
2457768 & $-6.8 \pm 14.5$ & no \\
2457772 & $12.1 \pm 14.6$ & no \\
2457779 & $-0.2 \pm 8.3$ & no \\
2457798 & $18.4 \pm 11.3$ & marginal \\
2457816 & $-24.3 \pm 8.4$ & marginal \\
2457817 & $-25.3 \pm 8.5$ & definite \\
\noalign{\smallskip} 
\hline
\end{tabular}
\tablefoot{Detection status refers to the Stokes V detection diagnosis \citep{1997MNRAS.291..658D}, and the longitudinal magnetic field was calculated following \citet{2022MNRAS.514.4300B}.}
\label{tab:espadons}
\end{table}

\subsection{Constraints on the long-term RV signal}  \label{subsec:second}

Besides the $\sim$390d signal that we classify as a Keplerian motion induced by a planetary companion, our RV data additionally show a long-term modulation with a timescale comparable to the baseline of the observations. We choose to fit a parabola to this modulation as the simplest model providing a good match to the data (Table\,\ref{table_lnZ}). However, this does not provide any information as to the nature of the observed RV variability. 

We can assume that the curvature in the RVs is due to a long-period companion. Following \citet{Kipping2011}, the approximate minimum-velocity semi-amplitude $K>7.5$ m s$^{-1}$ and measured RV acceleration $\ddot{\gamma}=-7.54^{+0.59}_{-0.61} \times 10^{-6}$\,m\,s$^{-1}$\,d$^{-2}$ can be combined to provide a lower limit on the orbital period $P_2>17.2\pm0.7$\,yr and minimum mass $M_2\sin{i} > 0.16\pm0.10$\,M$_\mathrm{Jup}$. We expand on this possibility using astrometric observations in the following section.

The possibility of a long-period companion was tested in our modelling. The posteriors of the fit with a Keplerian could be compatible with a sub-Saturn planet on a highly eccentric orbit, $e\sim0.5$ at 1.7\,au. However, the resulting period, 2\,268\,d, is comparable to the time baseline and well above the period calculated from the Nyquist frequency. Moreover, the eccentricity value seems to depend strongly on one single measurement at BJD 2458140 (Fig.\,\ref{fig:best_models1}), which casts doubt on the robustness of the solution. 

Giving the large scatter of 3\,m\,s$^{-1}$ of the uncorrelated residual RVs, the possibility for this modulation being of stellar origin seems high. As already explained, we considered a model with a GP and the QPC kernel (Sect.\,\ref{sec:RV}) to account for the possibility that stellar surface phenomena imprint the rotational period on the RVs, but no conclusive results were obtained. However, the variations of scatter of the RV residuals and of the H$\alpha$-index measurements, and the long-term periodicities found in some activity indicators and photometric time series, indicate a stellar influence on the data. However, this influence could only depend on episodic increases in the magnetic activity level of GJ\,1151. As it oscillates between more or less active phases, the star would imprint a larger or smaller scatter on the RV data, respectively.

\subsection{Constraints on companions from absolute astrometry}

Additional constraints on an outer companion can be provided based on information from the 320 observations from the {\it Gaia} DR3 archive. The renormalised unit weight error (RUWE) statistic for GJ\,1151 is 0.99, a value well below the threshold of 1.4 \citep[e.g.][]{Lindegren2018, Lindegren2021}. Therefore, a single-star model describes the data satisfactorily. We further followed the approach by e.g. \citet{Belokurov2020} and \citet{Penoyre2020} to investigate the range of orbital separations and companion masses that would induce excess astrometric residuals with respect to a single-star model, therefore producing larger RUWE values.

\begin{figure}[]
\centering
\includegraphics[width=\columnwidth]{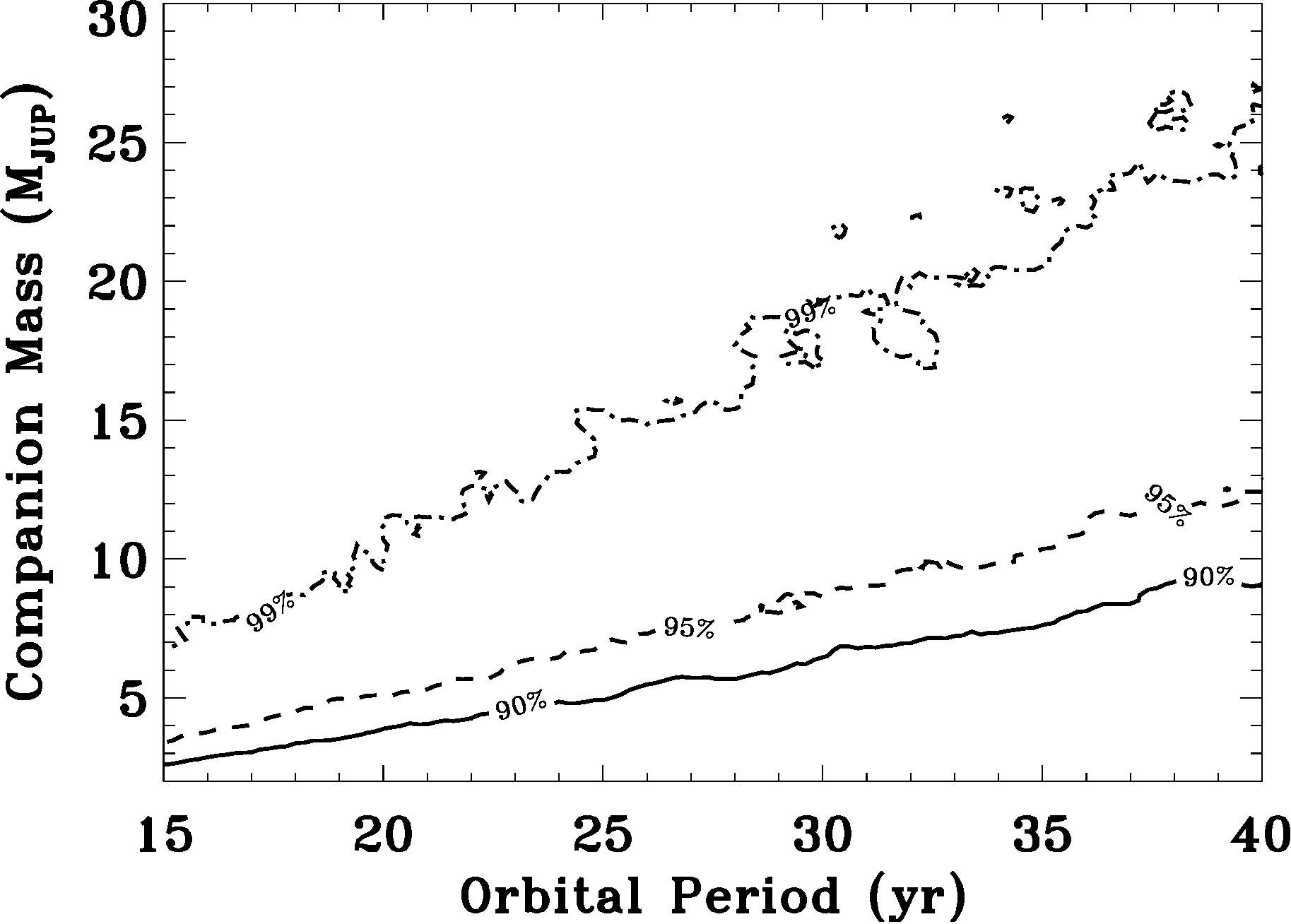}
\caption{\label{fig:GaiaRUWE_GJ1151} {\it Gaia} DR3 sensitivity to companions of given mass (in $M_{\rm Jup}$) as a function of orbital period (in yr) orbiting GJ\,1151. Solid, dashed, and dashed-dotted lines correspond to iso-probability curves for 90\%, 95\%, and 99\% probability of a companion of the given properties producing RUWE $>0.99$.}
\end{figure}

We generated synthetic {\it Gaia} observations of GJ\,1151 using the nominal astrometric parameters, and the values of observation times, scan angle, and along-scan parallax factors encompassing the mission time-span using the {\it Gaia} observation forecast tool\footnote{\url{https://gaia.esac.esa.int/gost/}}. We linearly added orbital motion effects due to companions with orbital periods in the range from 15 to 40\,yr and planetary mass in the range from 0.2 to 40\,M$_\mathrm{Jup}$, taking into account the lower limits on both parameters from the RV datasets, and assuming a primary mass of 0.16\,M$_\odot$ from Table\,\ref{table:GJ1151data}. One hundred random realisations of the remaining orbital elements were produced for each mass--period pair, all drawn from uniform distributions within their nominal intervals. We then perturbed the observations of each system with Gaussian measurement uncertainties appropriate for the case of a $G=11.7$\,mag star such as GJ\,1151 \citep[see e.g. Fig.\,3 in][]{Holl2022}. Finally, we performed linear fits to the synthetic data using a single-star model, and recorded the fraction of systems with RUWE exceeding 0.99 for each period--mass pair. Figure \ref{fig:GaiaRUWE_GJ1151} shows iso-probability contours in companion mass--period space corresponding to the fractions of systems with RUWE $>0.99$. Already for the minimum of a 17\,yr orbit, the planetary mass of a companion would be of $3 (8)$ M$_\mathrm{Jup}$ in order to produce a RUWE in excess of 0.99 with 90\% (99\%) probability. 

In summary, {\it Gaia} DR3 astrometry does not have sufficient sensitivity to probe for the presence of a giant outer companion of GJ\,1151 with true masses close to the lower limit set by the constraints from RVs. 

\subsection{Detection limits for a short-period planet}  \label{subsec:setlim}

\begin{figure}
\centering
\includegraphics[width=9 cm]{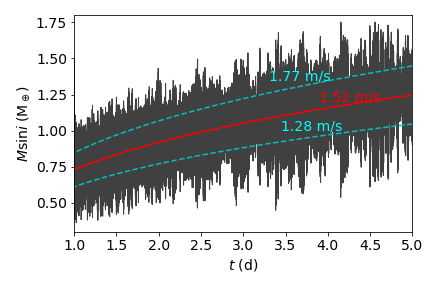}
 \caption{Detection limits of the RV residuals for the period of the assumed low-mass planet responsible for the LOFAR signal, after subtracting the best-fitting model consisting of the effect of GJ\,1151\,b and an additional quadratic polynomial. The black lines correspond to the minimum masses of the planetary signals as recovered by the simulations. The red line corresponds to the minimum mass of the average limiting semi-amplitude along with its standard deviation limits in blue.} \label{fig:detection_limits}
\end{figure}

The residuals from the best-fitting model produce a periodogram with no significant remaining signals (Fig.\,\ref{fig:best_models1}), although with a large scatter of 3\,m\,s$^{-1}$. For this reason, a yet undetected terrestrial planet compatible with the constraints from \citet{2020NatAs...4..577V} could be responsible for the LOFAR radio emission. Assuming that the residuals are uncorrelated, we calculate the detection limits of our data following \citet{2013A&A...549A.109B} and \citet{2009A&A...505..859Z}. To test this possibility, we inject circular planetary orbits with periods from 1 to 5\,d, with a linear sampling of (20\,000\,d)$^{-1}$ for 12 different phases. We start our trials with the semi-amplitude of the best-fit sine wave at every frequency. We iteratively increase the semi-amplitude of the injected planet in steps of 0.01 m\,s$^{-1}$ until it generates a significant signal in the periodogram of the residuals (FAP$<0.1\%$). The FAP is calculated using the approximate method by \citet{2008MNRAS.385.1279B}, which is computationally less expensive than the bootstrap method used before (see Sect.\,\ref{subsec:freq}). Therefore, we obtain the lower limit semi-amplitude for a significant signal to be detectable in our data given the noise level. Finally, we convert this value into an upper limit on the minimum mass of a planet to remain undetected. As we can see in Fig.\,\ref{fig:detection_limits}, we determine a limiting semi-amplitude of $K=1.52\pm0.25$\,m\,s$^{-1}$ for the studied period range, which translates into upper limits on the minimum mass of a planet from 0.73 to 1.25\,M$_\oplus$ for 1 to 5d orbits, respectively. We thereby only confirm our previously published value \citep{2021A&A...649L..12P} because of the increase in the RV residual scatter in the last observational season.

\subsection{LOFAR radio signal}  \label{subsec:radiotest}

The LOFAR radio emission detection by \citet{2020NatAs...4..577V} placed GJ\,1151 in the spotlight. The signal was found by cross-matching nearby stars ($d < 20$\,pc) from the {\it Gaia} database with the LOFAR Two-Metre Sky Survey \citep[LoTSS,][]{2019A&A...622A...1S} data release \citep[a similar following study,][expanded such detections to a further 19 M-dwarfs]{2021NatAs...5.1233C}. In particular, \citet{2020NatAs...4..577V} found a highly significant signal in one of the eight-hour pointings, during which the emission was detected for the full length of the observation. The emission was broadband, covering the entirety of the observed frequency range $120 - 167$\,MHz, with a nearly flat spectral shape. The emission was coherent and highly circularly polarised. All these features point to the electron cyclotron maser (ECM) mechanism, where a population of electrons group together in gyration phase and emit strongly. The resulting cut-off frequency, $\nu = 2.8\,\textbf{B}$[G] MHz, is a direct indication of the local magnetic field. The LoTSS upper frequency (167 MHz) then points to a local magnetic field of at least ${\bf B} \sim 60$\,G. However, we note that despite the multiple observations of GJ~1151 with the LOFAR instrument, the source has not been detected ever since (Vedantham, priv. comm.). 

With our current data and analysis, we can perform an assessment of the possible source of the low-frequency radio emission. We discuss four possible scenarios below.

{\it Stellar emission:} Radio emission from stars is associated with strong activity and magnetic fields stronger than the upper limit of 680\,G inferred by \citet{2022A&A...662A..41R}. It would be quite uncommon for the observed radio emission to be produced by GJ\,1151 itself. Also, the characteristics (duration, polarisation, bandwidth) of the observation do not fit into the standard emission processes seen on stars, whether of incoherent gyrosynchrotron origin, or from coherent plasma emission \citep{2020NatAs...4..577V}. Nevertheless, we cannot discard that the LOFAR detection is probing a yet unexplored domain of stellar radio emission at low frequencies, especially considering the apparent episodes of increased activity of the star. 

{\it Planetary emission:} The observed radio signal could be produced by auroral activity on the newly discovered planet GJ\,1151\,b. However, planets with sub-Jovian masses are not expected to possess strong magnetic fields \citep[e.g. ][, Table\,1]{2010SSRv..152..651S}. The dependence of the emitted frequencies on the magnetic field, which further depends on the planetary mass, suggests that a magnetic field ${\bf B} > 60$\,G would need a giant planet with a mass of at least $\sim 700 M_\oplus$ (according to the scaling law by \citealt{griessmeier07}; other scaling laws provide similarly large values). This mass is more than twice the mass of Jupiter. Agreement with our orbital solution and RV semi-amplitude would require a very small, very unlikely orbital inclination angle, with a nearly face-on configuration ($\sin{i} < 0.016, i<0.9$\,deg, probability $\sim 0.01$\,\%). Using the same approach as that described in Section 5.4, we find that, at a separation of 0.57 au, GJ 1151 b would produce a  larger RUWE than the one reported in the Gaia DR3 archive  with 99\% probability if its true mass is $\gtrsim 1$ M$_\mathrm{Jup}$, corresponding to $i\simeq2$ deg. The upper limit on the true mass of GJ 1151 b  from Gaia DR3 astrometry is not very informative, but it helps to narrowly rule out a true mass of 2 M$_\mathrm{Jup}$ for the companion, casting further doubt on the hypothesis of radio emission from the planet itself.

\begin{figure}
\centering
\includegraphics[width=9cm]{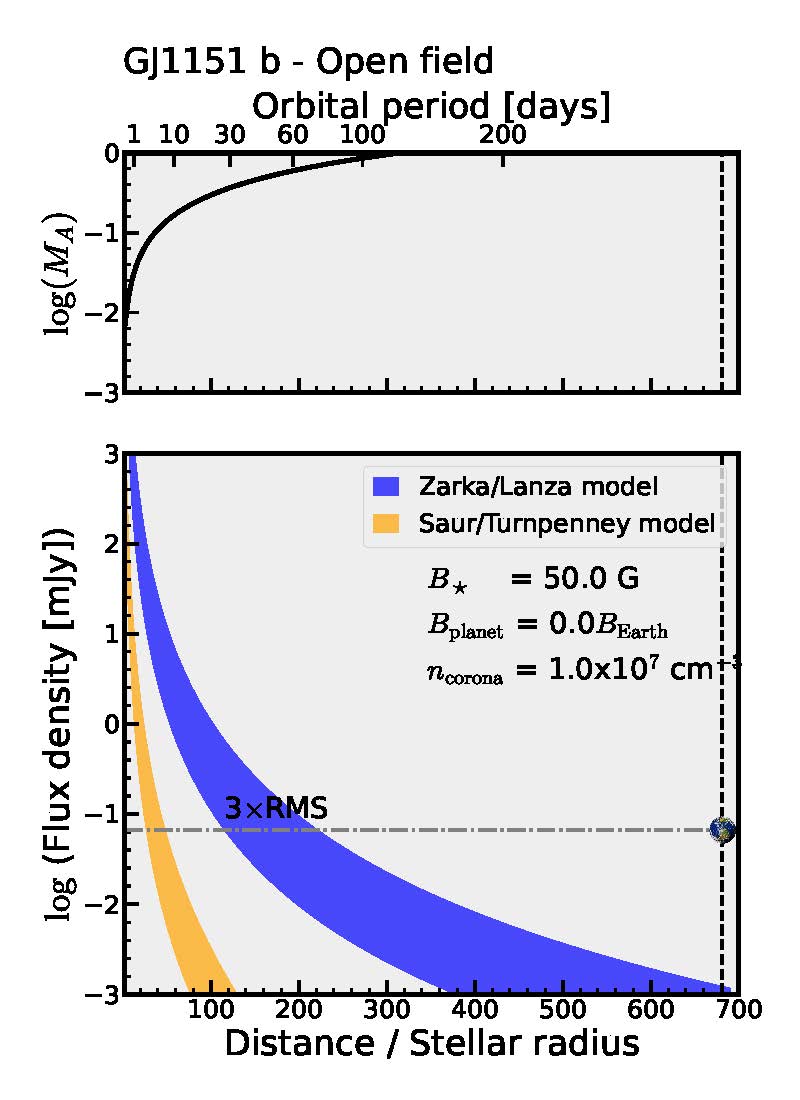}
 \caption{Expected flux density for auroral radio emission arising from a star--planet interaction in the system GJ\,1151 -- GJ\,1151\,b as a function of orbital distance. The interaction is clearly in the supra-Alfv\'enic regime (i.e. Alfv\'en Mach number $M_A = v_{\rm rel} \cdot v_{\rm Alfv}^{-1} > 1$; top panel) at the location of the planet (vertical dashed line). Therefore, star--planet interaction cannot account for the observed LOFAR radio emission. The planet in the bottom panel is drawn at around 0.1\,mJy in the y-axis, which corresponds to the 1-$\sigma$ level of the LOFAR radio observations.
 See main text for details.\label{fig:radio}}
\end{figure}

{\it Planet--star interaction:} GJ\,1151\,b could interact with its host star, giving rise to stellar auroral radio emission. We computed the radio emission arising from this star--planet interaction for both an open and a closed dipolar geometry, and for two different models of star-planet interaction: the Zarka-Lanza model \citep{zarka07,lanza09} and the Saur-Turnpenney model \citep{saur13,turnpenney18}. For illustration purposes, we show the results for the open magnetic field topologue  in Fig. \ref{fig:radio}. The bottom panel shows the predicted flux density as a function of orbital distance arising from a star--planet interaction involving GJ\,1151\,b for the case of an open magnetic field geometry of the star, following the prescriptions in \citet{2021A&A...645A..77P}, and assuming a 50\,G magnetic field at the local site where the cyclotron emission is produced, and an unmagnetised planet. This magnetic field corresponds to an electron--cyclotron frequency of 140\,MHz, which is the central observing frequency of the LOFAR observations. We note that the level of expected radio emission at the orbital position of GJ\,1151\,b is orders of magnitude below the typical detection sensitivity of LOFAR ($\sim 0.1$ mJy after a typical 4-8 hour run). We further note that for any realistic magnetic field of the planet, the expected flux density level remains well below any detection limit. The top panel shows the Alfv\'en Mach number as a function of orbital distance and orbital period. The interaction between planet GJ\,1151\,b and its host star happens in the supra-Alfv\'enic regime, meaning that energy and momentum cannot be transferred to the star. That is, for orbital periods above approximately 100\,days, all motion is in the supra-Alfv\'enic regime, and the flux densities drawn in the bottom panel do not apply, as the Poynting flux needed to feed the radio emission is never transferred to the star. Thus, star--planet interaction between GJ\,1151 and GJ\,1151\,b cannot account for the observed LOFAR radio emission. For further details, we refer to \citet{2021A&A...645A..77P}. 

{\it Inner planet:} A low-mass planet in a close orbit around its (weakly) magnetised host star (${\bf B} = 60$\,G, or 30\,G in the case of bright second harmonic emission) could induce the detected radio emission, as proposed initially by \citet{2020NatAs...4..577V}. As calculated in the previous section, the planet would have to have a minimum mass of below 1.2\,M$_\oplus$ and would therefore remain undetected given the properties of our dataset. 

\section{Conclusions}  \label{sec:con}

The M4.5 dwarf GJ\,1151 was monitored from 2016 to 2022 with the CARMENES, HARPS-N, and HPF instruments. The resulting datasets can be used to characterise both the planetary system of the star as well as the effects arising from stellar magnetic activity. 

We report the discovery of a planet compatible with sub-Neptune mass, $M \sin  i = 10.62^{+1.31}_{-1.47}$\,M$_\oplus$, in a circular orbit with a period of $P=389.7^{+5.4}_{-6.5}$\,d and a semi-major axis of $a=0.5714^{+0.0053}_{-0.0064}$\,au. We confirm the planetary nature of the radial-velocity signal by analysing spectroscopic activity diagnostics and also photometric variability of the star using archival data from MEarth and ASAS-SN, which have been collecting data on this target over the last 10 years. We update the rotation period of the star to $140\pm10$\,d, which is compatible with the upper limit of 2\,m\,s$^{-1}$ for $v \sin i$. We also analyse a large number of stellar indicators derived from our spectroscopic observations with CARMENES and HARPS-N. We confirm that GJ\,1151 is a star with a generally low magnetic activity level. However, the star shows episodes of larger magnetic activity where it induces variability in RVs, activity indices (e.g. H$\alpha$ emission), and photometry, and shows a detectable magnetic field with ESPaDOnS observations.

We find that it is highly unlikely that GJ 1151 b, by itself or by interacting with its host star, is responsible for the detected LOFAR radio emission. Also, it seems unlikely that an outer companion could be responsible, namely either a sub-Saturn on an eccentric orbit (as suggested by the RVs) or a Jupiter-like planet with $P > 17$\,yr (from the limits from {\it Gaia} observations). The Earth-mass planet in a close-in orbit proposed by \citet{2020NatAs...4..577V} remains a possible explanation for the radio signal, but  this possibility has a restricted parameter space  as it remains undetected in our data. Using the residuals from our best-fit model, we confirm the previous detection limits. Those translate into an upper limit for the minimum mass of a planet candidate at 0.73 to 1.25\,M$_\oplus$ for 1 to 5d orbits, respectively. However, we note that the fact that LOFAR has not detected GJ 1151 again does not exclude the possibility that the observed radio emission could be due to a low-mass planetary companion. To shed light on these different scenarios, a simultaneous, multi-wavelength radio campaign with different facilities and covering a time baseline of a few days (in order to follow a possible planet along its orbit) could confirm the previous LOFAR detection, and determine the frequency cut-off, which would give a direct measurement of the surface magnetic field.

Particularly surprising is the fact that the remaining RV residuals after the subtraction of both the sinusoidal and quadratic models are still of the order of 3\,m\,s$^{-1}$. This appears to be quite high for a star with such a low activity level given the reported experience from the CARMENES survey \citep{2022ribas}. The target will certainly continue to be observed in the future with the CARMENES instrument with the goal of better constraining the long-term modulations and the probably strong influence of the varying levels of stellar activity of GJ\,1151. The apparently episodic detection of a coherent, polarised, and broad-band radio emission, although not predicted for a generally inactive star such as GJ\,1151, could be the result of an unknown mechanism able to produce low-frequency emission in weakly magnetised stars, possibly compatible with at least some of the 19 signals detected in other {\it Gaia}-LOFAR matching searches \citep{2021NatAs...5.1233C}.

\begin{acknowledgements}
  CARMENES is an instrument at the Centro Astron\'omico Hispano en Andaluc\'ia (CAHA) at Calar Alto (Almer\'{\i}a, Spain), operated jointly by the Junta de Andaluc\'ia and the Instituto de Astrof\'isica de Andaluc\'ia (CSIC).
  
  CARMENES was funded by the Max-Planck-Gesellschaft (MPG), 
  the Consejo Superior de Investigaciones Cient\'{\i}ficas (CSIC),
  the Ministerio de Econom\'ia y Competitividad (MINECO) and the European Regional Development Fund (ERDF) through projects FICTS-2011-02, ICTS-2017-07-CAHA-4, and CAHA16-CE-3978, 
  and the members of the CARMENES Consortium 
  (Max-Planck-Institut f\"ur Astronomie,
  Instituto de Astrof\'{\i}sica de Andaluc\'{\i}a,
  Landessternwarte K\"onigstuhl,
  Institut de Ci\`encies de l'Espai,
  Institut f\"ur Astrophysik G\"ottingen,
  Universidad Complutense de Madrid,
  Th\"uringer Landessternwarte Tautenburg,
  Instituto de Astrof\'{\i}sica de Canarias,
  Hamburger Sternwarte,
  Centro de Astrobiolog\'{\i}a and
  Centro Astron\'omico Hispano-Alem\'an), 
  with additional contributions by the MINECO, 
  the Deutsche Forschungsgemeinschaft through the Major Research Instrumentation Programme and Research Unit FOR2544 ``Blue Planets around Red Stars'', 
  the Klaus Tschira Stiftung, 
  the states of Baden-W\"urttemberg and Niedersachsen, 
  and by the Junta de Andaluc\'{\i}a.
  
This work was partly based on observations made with the Italian {\it Telescopio Nazionale Galileo} (TNG) operated by the {Fundaci\'on Galileo Galilei} (FGG) of the {Istituto Nazionale di Astrofisica} (INAF) at the {Observatorio del Roque de los Muchachos} (La Palma, Canary Islands, Spain). 

We acknowledge financial support from the Agencia Estatal de Investigaci\'on of the Ministerio de Ciencia e Innovaci\'on MCIN/AEI/10.13039/501100011033 and the ERDF ``A way of making Europe'' through projects 
  PID2020-120375GB-I00,         
  PID2020-117404GB-C21,         
  PID2019-109522GB-C5[1:4],         
  PGC2018-098153-B-C33,         
  AYA2016-79425-C3-1,           
and the Centre of Excellence ``Severo Ochoa'' and ``Mar\'ia de Maeztu'' awards to the 
Institut de Ciències de l'Espai (CEX2020-001058-M)
Instituto de Astrof\'isica de Canarias (CEX2019-000920-S), Instituto de Astrof\'isica de Andaluc\'ia (SEV-2017-0709), and Centro de Astrobiolog\'ia (MDM-2017-0737);
the Generalitat de Catalunya/CERCA programme;
the European Union through a Marie Curie Action (101030103); 
the European Research Council (ERC) under the European Union's Horizon 2020 research and innovation programme (ERC Starting Grant IMAGINE, No. 948582); 
the DFG through priority program SPP 1992 ``Exploring the Diversity of Extrasolar Planets'' (JE 701/5-1); 
and the Italian Space Agency (ASI) in collaboration with the Istituto Nazionale di Astrofisica under contract 2018-24-HH.0.

We made use of data from the European Space Agency (ESA) mission {\it Gaia} (\url{https://www.cosmos.esa.int/gaia}), processed by the {\it Gaia} Data Processing and Analysis Consortium (DPAC, \url{https://www.cosmos.esa.int/web/gaia/dpac/consortium}). Funding for the DPAC has been provided by national institutions, in particular the institutions participating in the {\it Gaia} Multilateral Agreement. 

\end{acknowledgements}
\bibliographystyle{aa}
\bibliography{bibliography} 

\begin{thebibliography}{89}
\expandafter\ifx\csname natexlab\endcsname\relax\def\natexlab#1{#1}\fi

\bibitem[{{Affer} {et~al.}(2016){Affer}, {Micela}, {Damasso}, {Perger},
  {Ribas}, {Su{\'a}rez Mascare{\~n}o}, {Gonz{\'a}lez Hern{\'a}ndez}, {Rebolo},
  {Poretti}, {Maldonado}, {Leto}, {Pagano}, {Scandariato}, {Zanmar Sanchez},
  {Sozzetti}, {Bonomo}, {Malavolta}, {Morales}, {Rosich}, {Bignamini},
  {Gratton}, {Velasco}, {Cenadelli}, {Claudi}, {Cosentino}, {Desidera},
  {Giacobbe}, {Herrero}, {Lafarga}, {Lanza}, {Molinari}, \&
  {Piotto}}]{2016A&A...593A.117A}
{Affer}, L., {Micela}, G., {Damasso}, M., {et~al.} 2016, \aap, 593, A117

\bibitem[{{Anglada-Escud{\'e}} \& {Butler}(2012)}]{2012ApJS..200...15A}
{Anglada-Escud{\'e}}, G. \& {Butler}, R.~P. 2012, \apjs, 200, 15

\bibitem[{{Baglin} {et~al.}(2006){Baglin}, {Auvergne}, {Boisnard}, {Lam-Trong},
  {Barge}, {Catala}, {Deleuil}, {Michel}, \& {Weiss}}]{2006cosp...36.3749B}
{Baglin}, A., {Auvergne}, M., {Boisnard}, L., {et~al.} 2006, in 36th COSPAR
  Scientific Assembly, Vol.~36, 3749

\bibitem[{{Baluev}(2008)}]{2008MNRAS.385.1279B}
{Baluev}, R.~V. 2008, \mnras, 385, 1279

\bibitem[{{Baroch} {et~al.}(2020){Baroch}, {Morales}, {Ribas}, {Herrero},
  {Rosich}, {Perger}, {Anglada-Escud{\'e}}, {Reiners}, {Caballero},
  {Quirrenbach}, {Amado}, {Jeffers}, {Cifuentes}, {Passegger}, {Schweitzer},
  {Lafarga}, {Bauer}, {B{\'e}jar}, {Colom{\'e}}, {Cort{\'e}s-Contreras},
  {Dreizler}, {Galad{\'\i}-Enr{\'\i}quez}, {Hatzes}, {Henning}, {Kaminski},
  {K{\"u}rster}, {Montes}, {Rodr{\'\i}guez-L{\'o}pez}, \&
  {Zechmeister}}]{2020A&A...641A..69B}
{Baroch}, D., {Morales}, J.~C., {Ribas}, I., {et~al.} 2020, \aap, 641, A69

\bibitem[{{Belokurov} {et~al.}(2020){Belokurov}, {Penoyre}, {Oh}, {Iorio},
  {Hodgkin}, {Evans}, {Everall}, {Koposov}, {Tout}, {Izzard}, {Clarke}, \&
  {Brown}}]{Belokurov2020}
{Belokurov}, V., {Penoyre}, Z., {Oh}, S., {et~al.} 2020, \mnras, 496, 1922

\bibitem[{{Bonfils} {et~al.}(2013){Bonfils}, {Delfosse}, {Udry}, {Forveille},
  {Mayor}, {Perrier}, {Bouchy}, {Gillon}, {Lovis}, {Pepe}, {Queloz}, {Santos},
  {S{\'e}gransan}, \& {Bertaux}}]{2013A&A...549A.109B}
{Bonfils}, X., {Delfosse}, X., {Udry}, S., {et~al.} 2013, \aap, 549, A109

\bibitem[{{Borucki} {et~al.}(2011){Borucki}, {Koch}, {Basri}, {Batalha},
  {Brown}, {Bryson}, {Caldwell}, {Christensen-Dalsgaard}, {Cochran}, {DeVore},
  {Dunham}, {Gautier}, {Geary}, {Gilliland}, {Gould}, {Howell}, {Jenkins},
  {Latham}, {Lissauer}, {Marcy}, {Rowe}, {Sasselov}, {Boss}, {Charbonneau},
  {Ciardi}, {Doyle}, {Dupree}, {Ford}, {Fortney}, {Holman}, {Seager},
  {Steffen}, {Tarter}, {Welsh}, {Allen}, {Buchhave}, {Christiansen}, {Clarke},
  {Das}, {D{\'e}sert}, {Endl}, {Fabrycky}, {Fressin}, {Haas}, {Horch},
  {Howard}, {Isaacson}, {Kjeldsen}, {Kolodziejczak}, {Kulesa}, {Li}, {Lucas},
  {Machalek}, {McCarthy}, {MacQueen}, {Meibom}, {Miquel}, {Prsa}, {Quinn},
  {Quintana}, {Ragozzine}, {Sherry}, {Shporer}, {Tenenbaum}, {Torres},
  {Twicken}, {Van Cleve}, {Walkowicz}, {Witteborn}, \&
  {Still}}]{2011ApJ...736...19B}
{Borucki}, W.~J., {Koch}, D.~G., {Basri}, G., {et~al.} 2011, \apj, 736, 19

\bibitem[{{Brown} {et~al.}(2022){Brown}, {Jeffers}, {Marsden}, {Morin}, {Boro
  Saikia}, {Petit}, {Jardine}, {See}, {Vidotto}, {Mengel}, {Dahlkemper}, \&
  {the BCool Collaboration}}]{2022MNRAS.514.4300B}
{Brown}, E.~L., {Jeffers}, S.~V., {Marsden}, S.~C., {et~al.} 2022, \mnras, 514,
  4300

\bibitem[{{Caballero} {et~al.}(2013){Caballero}, {Cort{\'e}s-Contreras},
  {L{\'o}pez-Santiago}, {Alonso-Floriano}, {Klutsch}, {Montes}, {Morales},
  {Mundt}, {Ribas}, {Reiners}, {Quirrenbach}, {Amado}, {CARMENES Consortium},
  \& {del Burgo}}]{2013hsa7.conf..645C}
{Caballero}, J.~A., {Cort{\'e}s-Contreras}, M., {L{\'o}pez-Santiago}, J.,
  {et~al.} 2013, in Highlights of Spanish Astrophysics VII, ed. J.~C.
  {Guirado}, L.~M. {Lara}, V.~{Quilis}, \& J.~{Gorgas}, 645--645

\bibitem[{{Caballero} {et~al.}(2016){Caballero}, {Gu{\`a}rdia}, {L{\'o}pez del
  Fresno}, {Zechmeister}, {de Juan}, {Alonso-Floriano}, {Amado}, {Colom{\'e}},
  {Cort{\'e}s-Contreras}, {Garc{\'\i}a-Piquer}, {Gesa}, {de Guindos}, {Hagen},
  {Helmling}, {Hern{\'a}ndez Casta{\~n}o}, {K{\"u}rster}, {L{\'o}pez-Santiago},
  {Montes}, {Morales Mu{\~n}oz}, {Pavlov}, {Quirrenbach}, {Reiners}, {Ribas},
  {Seifert}, \& {Solano}}]{2016SPIE.9910E..0EC}
{Caballero}, J.~A., {Gu{\`a}rdia}, J., {L{\'o}pez del Fresno}, M., {et~al.}
  2016, in Society of Photo-Optical Instrumentation Engineers (SPIE) Conference
  Series, Vol. 9910, Observatory Operations: Strategies, Processes, and Systems
  VI, ed. A.~B. {Peck}, R.~L. {Seaman}, \& C.~R. {Benn}, 99100E

\bibitem[{{Callingham} {et~al.}(2021){Callingham}, {Vedantham}, {Shimwell},
  {Pope}, {Davis}, {Best}, {Hardcastle}, {R{\"o}ttgering}, {Sabater}, {Tasse},
  {van Weeren}, {Williams}, {Zarka}, {de Gasperin}, \&
  {Drabent}}]{2021NatAs...5.1233C}
{Callingham}, J.~R., {Vedantham}, H.~K., {Shimwell}, T.~W., {et~al.} 2021,
  Nature Astronomy, 5, 1233

\bibitem[{{Charbonneau} {et~al.}(2008){Charbonneau}, {Irwin}, {Nutzman}, \&
  {Falco}}]{2008AAS...212.4402C}
{Charbonneau}, D., {Irwin}, J., {Nutzman}, P., \& {Falco}, E.~E. 2008, in
  American Astronomical Society Meeting Abstracts \#212, 44.02

\bibitem[{{Cifuentes} {et~al.}(2020){Cifuentes}, {Caballero},
  {Cort{\'e}s-Contreras}, {Montes}, {Abell{\'a}n}, {Dorda}, {Holgado},
  {Zapatero Osorio}, {Morales}, {Amado}, {Passegger}, {Quirrenbach}, {Reiners},
  {Ribas}, {Sanz-Forcada}, {Schweitzer}, {Seifert}, \&
  {Solano}}]{2020A&A...642A.115C}
{Cifuentes}, C., {Caballero}, J.~A., {Cort{\'e}s-Contreras}, M., {et~al.} 2020,
  \aap, 642, A115

\bibitem[{{Cosentino} {et~al.}(2012){Cosentino}, {Lovis}, {Pepe}, {Collier
  Cameron}, {Latham}, {Molinari}, {Udry}, {Bezawada}, {Black}, {Born},
  {Buchschacher}, {Charbonneau}, {Figueira}, {Fleury}, {Galli}, {Gallie},
  {Gao}, {Ghedina}, {Gonzalez}, {Gonzalez}, {Guerra}, {Henry}, {Horne},
  {Hughes}, {Kelly}, {Lodi}, {Lunney}, {Maire}, {Mayor}, {Micela}, {Ordway},
  {Peacock}, {Phillips}, {Piotto}, {Pollacco}, {Queloz}, {Rice}, {Riverol},
  {Riverol}, {San Juan}, {Sasselov}, {Segransan}, {Sozzetti}, {Sosnowska},
  {Stobie}, {Szentgyorgyi}, {Vick}, \& {Weber}}]{2012SPIE.8446E..1VC}
{Cosentino}, R., {Lovis}, C., {Pepe}, F., {et~al.} 2012, in Society of
  Photo-Optical Instrumentation Engineers (SPIE) Conference Series, Vol. 8446,
  Ground-based and Airborne Instrumentation for Astronomy IV, ed. I.~S.
  {McLean}, S.~K. {Ramsay}, \& H.~{Takami}, 84461V

\bibitem[{{Courcol} {et~al.}(2015){Courcol}, {Bouchy}, {Pepe}, {Santerne},
  {Delfosse}, {Arnold}, {Astudillo-Defru}, {Boisse}, {Bonfils}, {Borgniet},
  {Bourrier}, {Cabrera}, {Deleuil}, {Demangeon}, {D{\'\i}az}, {Ehrenreich},
  {Forveille}, {H{\'e}brard}, {Lagrange}, {Montagnier}, {Moutou}, {Rey},
  {Santos}, {S{\'e}gransan}, {Udry}, \& {Wilson}}]{2015A&A...581A..38C}
{Courcol}, B., {Bouchy}, F., {Pepe}, F., {et~al.} 2015, \aap, 581, A38

\bibitem[{{Cutri} {et~al.}(2003){Cutri}, {Skrutskie}, {van Dyk}, {Beichman},
  {Carpenter}, {Chester}, {Cambresy}, {Evans}, {Fowler}, {Gizis}, {Howard},
  {Huchra}, {Jarrett}, {Kopan}, {Kirkpatrick}, {Light}, {Marsh}, {McCallon},
  {Schneider}, {Stiening}, {Sykes}, {Weinberg}, {Wheaton}, {Wheelock}, \&
  {Zacarias}}]{2003yCat.2246....0C}
{Cutri}, R.~M., {Skrutskie}, M.~F., {van Dyk}, S., {et~al.} 2003, VizieR Online
  Data Catalog, II/246

\bibitem[{{Czesla} {et~al.}(2022){Czesla}, {Lamp{\'o}n}, {Sanz-Forcada},
  {Garc{\'\i}a Mu{\~n}oz}, {L{\'o}pez-Puertas}, {Nortmann}, {Yan}, {Nagel},
  {Yan}, {Schmitt}, {Aceituno}, {Amado}, {Caballero}, {Casasayas-Barris},
  {Henning}, {Khalafinejad}, {Molaverdikhani}, {Montes}, {Pall{\'e}},
  {Reiners}, {Schneider}, {Ribas}, {Quirrenbach}, {Zapatero Osorio}, \&
  {Zechmeister}}]{2022A&A...657A...6C}
{Czesla}, S., {Lamp{\'o}n}, M., {Sanz-Forcada}, J., {et~al.} 2022, \aap, 657,
  A6

\bibitem[{{D{\'\i}ez Alonso} {et~al.}(2019){D{\'\i}ez Alonso}, {Caballero},
  {Montes}, {de Cos Juez}, {Dreizler}, {Dubois}, {Jeffers}, {Lalitha}, {Naves},
  {Reiners}, {Ribas}, {Vanaverbeke}, {Amado}, {B{\'e}jar},
  {Cort{\'e}s-Contreras}, {Herrero}, {Hidalgo}, {K{\"u}rster}, {Logie},
  {Quirrenbach}, {Rau}, {Seifert}, {Sch{\"o}fer}, \&
  {Tal-Or}}]{2019A&A...621A.126D}
{D{\'\i}ez Alonso}, E., {Caballero}, J.~A., {Montes}, D., {et~al.} 2019, \aap,
  621, A126

\bibitem[{{Donati}(2003)}]{2003ASPC..307...41D}
{Donati}, J.~F. 2003, in Astronomical Society of the Pacific Conference Series,
  Vol. 307, Solar Polarization, ed. J.~{Trujillo-Bueno} \& J.~{Sanchez
  Almeida}, 41

\bibitem[{{Donati} {et~al.}(1997){Donati}, {Semel}, {Carter}, {Rees}, \&
  {Collier Cameron}}]{1997MNRAS.291..658D}
{Donati}, J.~F., {Semel}, M., {Carter}, B.~D., {Rees}, D.~E., \& {Collier
  Cameron}, A. 1997, \mnras, 291, 658

\bibitem[{{Dumusque} {et~al.}(2015){Dumusque}, {Pepe}, {Lovis}, \&
  {Latham}}]{2015ApJ...808..171D}
{Dumusque}, X., {Pepe}, F., {Lovis}, C., \& {Latham}, D.~W. 2015, \apj, 808,
  171

\bibitem[{Efron \& Tibshirani(1985)}]{efron1985bootstrap}
Efron, B. \& Tibshirani, R. 1985, Behaviormetrika, 12, 1

\bibitem[{{Espinoza} {et~al.}(2019){Espinoza}, {Kossakowski}, \&
  {Brahm}}]{2019MNRAS.490.2262E}
{Espinoza}, N., {Kossakowski}, D., \& {Brahm}, R. 2019, \mnras, 490, 2262

\bibitem[{{Foster} {et~al.}(2020){Foster}, {Poppenhaeger},
  {Alvarado-G{\'o}mez}, \& {Schmitt}}]{2020MNRAS.497.1015F}
{Foster}, G., {Poppenhaeger}, K., {Alvarado-G{\'o}mez}, J.~D., \& {Schmitt},
  J.~H.~M.~M. 2020, \mnras, 497, 1015

\bibitem[{{Gaia Collaboration} {et~al.}(2021){Gaia Collaboration}, {Brown},
  {Vallenari}, {Prusti}, {de Bruijne}, {Babusiaux}, {Biermann}, {Creevey},
  {Evans}, {Eyer}, {Hutton}, {Jansen}, {Jordi}, {Klioner}, {Lammers},
  {Lindegren}, {Luri}, {Mignard}, {Panem}, {Pourbaix}, {Randich}, {Sartoretti},
  {Soubiran}, {Walton}, {Arenou}, {Bailer-Jones}, {Bastian}, {Cropper},
  {Drimmel}, {Katz}, {Lattanzi}, {van Leeuwen}, {Bakker}, {Cacciari},
  {Casta{\~n}eda}, {De Angeli}, {Ducourant}, {Fabricius}, {Fouesneau},
  {Fr{\'e}mat}, {Guerra}, {Guerrier}, {Guiraud}, {Jean-Antoine Piccolo},
  {Masana}, {Messineo}, {Mowlavi}, {Nicolas}, {Nienartowicz}, {Pailler},
  {Panuzzo}, {Riclet}, {Roux}, {Seabroke}, {Sordo}, {Tanga}, {Th{\'e}venin},
  {Gracia-Abril}, {Portell}, {Teyssier}, {Altmann}, {Andrae}, {Bellas-Velidis},
  {Benson}, {Berthier}, {Blomme}, {Brugaletta}, {Burgess}, {Busso}, {Carry},
  {Cellino}, {Cheek}, {Clementini}, {Damerdji}, {Davidson}, {Delchambre},
  {Dell'Oro}, {Fern{\'a}ndez-Hern{\'a}ndez}, {Galluccio}, {Garc{\'\i}a-Lario},
  {Garcia-Reinaldos}, {Gonz{\'a}lez-N{\'u}{\~n}ez}, {Gosset}, {Haigron},
  {Halbwachs}, {Hambly}, {Harrison}, {Hatzidimitriou}, {Heiter},
  {Hern{\'a}ndez}, {Hestroffer}, {Hodgkin}, {Holl}, {Jan{\ss}en}, {Jevardat de
  Fombelle}, {Jordan}, {Krone-Martins}, {Lanzafame}, {L{\"o}ffler}, {Lorca},
  {Manteiga}, {Marchal}, {Marrese}, {Moitinho}, {Mora}, {Muinonen}, {Osborne},
  {Pancino}, {Pauwels}, {Petit}, {Recio-Blanco}, {Richards}, {Riello},
  {Rimoldini}, {Robin}, {Roegiers}, {Rybizki}, {Sarro}, {Siopis}, {Smith},
  {Sozzetti}, {Ulla}, {Utrilla}, {van Leeuwen}, {van Reeven}, {Abbas}, {Abreu
  Aramburu}, {Accart}, {Aerts}, {Aguado}, {Ajaj}, {Altavilla}, {{\'A}lvarez},
  {{\'A}lvarez Cid-Fuentes}, {Alves}, {Anderson}, {Anglada Varela}, {Antoja},
  {Audard}, {Baines}, {Baker}, {Balaguer-N{\'u}{\~n}ez}, {Balbinot}, {Balog},
  {Barache}, {Barbato}, {Barros}, {Barstow}, {Bartolom{\'e}}, {Bassilana},
  {Bauchet}, {Baudesson-Stella}, {Becciani}, {Bellazzini}, {Bernet}, {Bertone},
  {Bianchi}, {Blanco-Cuaresma}, {Boch}, {Bombrun}, {Bossini}, {Bouquillon},
  {Bragaglia}, {Bramante}, {Breedt}, {Bressan}, {Brouillet}, {Bucciarelli},
  {Burlacu}, {Busonero}, {Butkevich}, {Buzzi}, {Caffau}, {Cancelliere},
  {C{\'a}novas}, {Cantat-Gaudin}, {Carballo}, {Carlucci}, {Carnerero},
  {Carrasco}, {Casamiquela}, {Castellani}, {Castro-Ginard}, {Castro Sampol},
  {Chaoul}, {Charlot}, {Chemin}, {Chiavassa}, {Cioni}, {Comoretto}, {Cooper},
  {Cornez}, {Cowell}, {Crifo}, {Crosta}, {Crowley}, {Dafonte}, {Dapergolas},
  {David}, {David}, {de Laverny}, {De Luise}, {De March}, {De Ridder}, {de
  Souza}, {de Teodoro}, {de Torres}, {del Peloso}, {del Pozo}, {Delbo},
  {Delgado}, {Delgado}, {Delisle}, {Di Matteo}, {Diakite}, {Diener},
  {Distefano}, {Dolding}, {Eappachen}, {Edvardsson}, {Enke}, {Esquej}, {Fabre},
  {Fabrizio}, {Faigler}, {Fedorets}, {Fernique}, {Fienga}, {Figueras},
  {Fouron}, {Fragkoudi}, {Fraile}, {Franke}, {Gai}, {Garabato},
  {Garcia-Gutierrez}, {Garc{\'\i}a-Torres}, {Garofalo}, {Gavras}, {Gerlach},
  {Geyer}, {Giacobbe}, {Gilmore}, {Girona}, {Giuffrida}, {Gomel}, {Gomez},
  {Gonzalez-Santamaria}, {Gonz{\'a}lez-Vidal}, {Granvik},
  {Guti{\'e}rrez-S{\'a}nchez}, {Guy}, {Hauser}, {Haywood}, {Helmi}, {Hidalgo},
  {Hilger}, {H{\l}adczuk}, {Hobbs}, {Holland}, {Huckle}, {Jasniewicz},
  {Jonker}, {Juaristi Campillo}, {Julbe}, {Karbevska}, {Kervella}, {Khanna},
  {Kochoska}, {Kontizas}, {Kordopatis}, {Korn}, {Kostrzewa-Rutkowska},
  {Kruszy{\'n}ska}, {Lambert}, {Lanza}, {Lasne}, {Le Campion}, {Le Fustec},
  {Lebreton}, {Lebzelter}, {Leccia}, {Leclerc}, {Lecoeur-Taibi}, {Liao},
  {Licata}, {Lindstr{\o}m}, {Lister}, {Livanou}, {Lobel}, {Madrero Pardo},
  {Managau}, {Mann}, {Marchant}, {Marconi}, {Marcos Santos}, {Marinoni},
  {Marocco}, {Marshall}, {Martin Polo}, {Mart{\'\i}n-Fleitas}, {Masip},
  {Massari}, {Mastrobuono-Battisti}, {Mazeh}, {McMillan}, {Messina},
  {Michalik}, {Millar}, {Mints}, {Molina}, {Molinaro}, {Moln{\'a}r},
  {Montegriffo}, {Mor}, {Morbidelli}, {Morel}, {Morris}, {Mulone}, {Munoz},
  {Muraveva}, {Murphy}, {Musella}, {Noval}, {Ord{\'e}novic}, {Orr{\`u}},
  {Osinde}, {Pagani}, {Pagano}, {Palaversa}, {Palicio}, {Panahi}, {Pawlak},
  {Pe{\~n}alosa Esteller}, {Penttil{\"a}}, {Piersimoni}, {Pineau}, {Plachy},
  {Plum}, {Poggio}, {Poretti}, {Poujoulet}, {Pr{\v{s}}a}, {Pulone}, {Racero},
  {Ragaini}, {Rainer}, {Raiteri}, {Rambaux}, {Ramos}, {Ramos-Lerate}, {Re
  Fiorentin}, {Regibo}, {Reyl{\'e}}, {Ripepi}, {Riva}, {Rixon}, {Robichon},
  {Robin}, {Roelens}, {Rohrbasser}, {Romero-G{\'o}mez}, {Rowell}, {Royer},
  {Rybicki}, {Sadowski}, {Sagrist{\`a} Sell{\'e}s}, {Sahlmann}, {Salgado},
  {Salguero}, {Samaras}, {Sanchez Gimenez}, {Sanna}, {Santove{\~n}a},
  {Sarasso}, {Schultheis}, {Sciacca}, {Segol}, {Segovia}, {S{\'e}gransan},
  {Semeux}, {Shahaf}, {Siddiqui}, {Siebert}, {Siltala}, {Slezak}, {Smart},
  {Solano}, {Solitro}, {Souami}, {Souchay}, {Spagna}, {Spoto}, {Steele},
  {Steidelm{\"u}ller}, {Stephenson}, {S{\"u}veges}, {Szabados}, {Szegedi-Elek},
  {Taris}, {Tauran}, {Taylor}, {Teixeira}, {Thuillot}, {Tonello}, {Torra},
  {Torra}, {Turon}, {Unger}, {Vaillant}, {van Dillen}, {Vanel}, {Vecchiato},
  {Viala}, {Vicente}, {Voutsinas}, {Weiler}, {Wevers}, {Wyrzykowski}, {Yoldas},
  {Yvard}, {Zhao}, {Zorec}, {Zucker}, {Zurbach}, \&
  {Zwitter}}]{2021A&A...649A...1G}
{Gaia Collaboration}, {Brown}, A.~G.~A., {Vallenari}, A., {et~al.} 2021, \aap,
  649, A1

\bibitem[{{Giclas} {et~al.}(1971){Giclas}, {Burnham}, \&
  {Thomas}}]{1971lpms.book.....G}
{Giclas}, H.~L., {Burnham}, R., \& {Thomas}, N.~G. 1971, {Lowell proper motion
  survey Northern Hemisphere. The G numbered stars. 8991 stars fainter than
  magnitude 8 with motions > 0''.26/year}

\bibitem[{{Gliese} \& {Jahrei{\ss}}(1979)}]{1979A&AS...38..423G}
{Gliese}, W. \& {Jahrei{\ss}}, H. 1979, \aaps, 38, 423

\bibitem[{{Grie{\ss}meier} {et~al.}(2007){Grie{\ss}meier}, {Zarka}, \&
  {Spreeuw}}]{griessmeier07}
{Grie{\ss}meier}, J.~M., {Zarka}, P., \& {Spreeuw}, H. 2007, \aap, 475, 359

\bibitem[{{Hallinan} {et~al.}(2015){Hallinan}, {Littlefair}, {Cotter},
  {Bourke}, {Harding}, {Pineda}, {Butler}, {Golden}, {Basri}, {Doyle}, {Kao},
  {Berdyugina}, {Kuznetsov}, {Rupen}, \& {Antonova}}]{2015Natur.523..568H}
{Hallinan}, G., {Littlefair}, S.~P., {Cotter}, G., {et~al.} 2015, \nat, 523,
  568

\bibitem[{{Hawley} {et~al.}(1996){Hawley}, {Gizis}, \&
  {Reid}}]{1996AJ....112.2799H}
{Hawley}, S.~L., {Gizis}, J.~E., \& {Reid}, I.~N. 1996, \aj, 112, 2799

\bibitem[{{Holl} {et~al.}(2022){Holl}, {Sozzetti}, {Sahlmann}, {Giacobbe},
  {S{\'e}gransan}, {Unger}, {Delisle}, {Barbato}, {Lattanzi}, {Morbidelli}, \&
  {Sosnowska}}]{Holl2022}
{Holl}, B., {Sozzetti}, A., {Sahlmann}, J., {et~al.} 2022, arXiv e-prints,
  arXiv:2206.05439

\bibitem[{{Jeffers} {et~al.}(2018){Jeffers}, {Sch{\"o}fer}, {Lamert},
  {Reiners}, {Montes}, {Caballero}, {Cort{\'e}s-Contreras}, {Marvin},
  {Passegger}, {Zechmeister}, {Quirrenbach}, {Alonso-Floriano}, {Amado},
  {Bauer}, {Casal}, {Diez Alonso}, {Herrero}, {Morales}, {Mundt}, {Ribas}, \&
  {Sarmiento}}]{2018A&A...614A..76J}
{Jeffers}, S.~V., {Sch{\"o}fer}, P., {Lamert}, A., {et~al.} 2018, \aap, 614,
  A76

\bibitem[{{Jones} {et~al.}(2017){Jones}, {Tuomi}, {Anglada-Escudé}, {Feng}, \&
  {Butler} R.~P.and~{Vogt}}]{2017AAS...22940307J}
{Jones}, H., {Tuomi}, M., {Anglada-Escudé}, G., {Feng}, F., \& {Butler}
  R.~P.and~{Vogt}, S. 2017, in American Astronomical Society Meeting Abstracts,
  Vol. 229, American Astronomical Society Meeting Abstracts \#229, 403.07

\bibitem[{{Kipping} {et~al.}(2011){Kipping}, {Hartman}, {Bakos}, {Torres},
  {Latham}, {Bayliss}, {Kiss}, {Sato}, {B{\'e}ky}, {Kov{\'a}cs}, {Quinn},
  {Buchhave}, {Andersen}, {Marcy}, {Howard}, {Fischer}, {Johnson}, {Noyes},
  {Sasselov}, {Stefanik}, {L{\'a}z{\'a}r}, {Papp}, {S{\'a}ri}, \&
  {F{\H{u}}r{\'e}sz}}]{Kipping2011}
{Kipping}, D.~M., {Hartman}, J., {Bakos}, G.~{\'A}., {et~al.} 2011, \aj, 142,
  95

\bibitem[{{Klein} {et~al.}(2022){Klein}, {Zicher}, {Kavanagh}, {Nielsen},
  {Aigrain}, {Vidotto}, {Barrag{\'a}n}, {Strugarek}, {Nicholson}, {Donati}, \&
  {Bouvier}}]{2022MNRAS.512.5067K}
{Klein}, B., {Zicher}, N., {Kavanagh}, R.~D., {et~al.} 2022, \mnras, 512, 5067

\bibitem[{{Kochanek} {et~al.}(2017){Kochanek}, {Shappee}, {Stanek}, {Holoien},
  {Thompson}, {Prieto}, {Dong}, {Shields}, {Will}, {Britt}, {Perzanowski}, \&
  {Pojma{\'n}ski}}]{2017PASP..129j4502K}
{Kochanek}, C.~S., {Shappee}, B.~J., {Stanek}, K.~Z., {et~al.} 2017, \pasp,
  129, 104502

\bibitem[{{Kopparapu} {et~al.}(2013){Kopparapu}, {Ramirez}, {Kasting}, {Eymet},
  {Robinson}, {Mahadevan}, {Terrien}, {Domagal-Goldman}, {Meadows}, \&
  {Deshpande}}]{2013ApJ...765..131K}
{Kopparapu}, R.~K., {Ramirez}, R., {Kasting}, J.~F., {et~al.} 2013, \apj, 765,
  131

\bibitem[{{K{\"u}rster} {et~al.}(2009){K{\"u}rster}, {Zechmeister}, {Endl}, \&
  {Meyer}}]{2009Msngr.136...39K}
{K{\"u}rster}, M., {Zechmeister}, M., {Endl}, M., \& {Meyer}, E. 2009, The
  Messenger, 136, 39

\bibitem[{{Lafarga} {et~al.}(2020){Lafarga}, {Ribas}, {Lovis}, {Perger},
  {Zechmeister}, {Bauer}, {K{\"u}rster}, {Cort{\'e}s-Contreras}, {Morales},
  {Herrero}, {Rosich}, {Baroch}, {Reiners}, {Caballero}, {Quirrenbach},
  {Amado}, {Alacid}, {B{\'e}jar}, {Dreizler}, {Hatzes}, {Henning}, {Jeffers},
  {Kaminski}, {Montes}, {Pedraz}, {Rodr{\'\i}guez-L{\'o}pez}, \&
  {Schmitt}}]{2020A&A...636A..36L}
{Lafarga}, M., {Ribas}, I., {Lovis}, C., {et~al.} 2020, \aap, 636, A36

\bibitem[{{Lafarga} {et~al.}(2021){Lafarga}, {Ribas}, {Reiners}, {Quirrenbach},
  {Amado}, {Caballero}, {Azzaro}, {B{\'e}jar}, {Cort{\'e}s-Contreras},
  {Dreizler}, {Hatzes}, {Henning}, {Jeffers}, {Kaminski}, {K{\"u}rster},
  {Montes}, {Morales}, {Oshagh}, {Rodr{\'\i}guez-L{\'o}pez}, {Sch{\"o}fer},
  {Schweitzer}, \& {Zechmeister}}]{2021A&A...652A..28L}
{Lafarga}, M., {Ribas}, I., {Reiners}, A., {et~al.} 2021, \aap, 652, A28

\bibitem[{{Lanza}(2009)}]{lanza09}
{Lanza}, A.~F. 2009, \aap, 505, 339

\bibitem[{{Lindegren} {et~al.}(2018){Lindegren}, {Hern{\'a}ndez}, {Bombrun},
  {Klioner}, {Bastian}, {Ramos-Lerate}, {de Torres}, {Steidelm{\"u}ller},
  {Stephenson}, {Hobbs}, {Lammers}, {Biermann}, {Geyer}, {Hilger}, {Michalik},
  {Stampa}, {McMillan}, {Casta{\~n}eda}, {Clotet}, {Comoretto}, {Davidson},
  {Fabricius}, {Gracia}, {Hambly}, {Hutton}, {Mora}, {Portell}, {van Leeuwen},
  {Abbas}, {Abreu}, {Altmann}, {Andrei}, {Anglada}, {Balaguer-N{\'u}{\~n}ez},
  {Barache}, {Becciani}, {Bertone}, {Bianchi}, {Bouquillon}, {Bourda},
  {Br{\"u}semeister}, {Bucciarelli}, {Busonero}, {Buzzi}, {Cancelliere},
  {Carlucci}, {Charlot}, {Cheek}, {Crosta}, {Crowley}, {de Bruijne}, {de
  Felice}, {Drimmel}, {Esquej}, {Fienga}, {Fraile}, {Gai}, {Garralda},
  {Gonz{\'a}lez-Vidal}, {Guerra}, {Hauser}, {Hofmann}, {Holl}, {Jordan},
  {Lattanzi}, {Lenhardt}, {Liao}, {Licata}, {Lister}, {L{\"o}ffler},
  {Marchant}, {Martin-Fleitas}, {Messineo}, {Mignard}, {Morbidelli}, {Poggio},
  {Riva}, {Rowell}, {Salguero}, {Sarasso}, {Sciacca}, {Siddiqui}, {Smart},
  {Spagna}, {Steele}, {Taris}, {Torra}, {van Elteren}, {van Reeven}, \&
  {Vecchiato}}]{Lindegren2018}
{Lindegren}, L., {Hern{\'a}ndez}, J., {Bombrun}, A., {et~al.} 2018, \aap, 616,
  A2

\bibitem[{{Lindegren} {et~al.}(2021){Lindegren}, {Klioner}, {Hern{\'a}ndez},
  {Bombrun}, {Ramos-Lerate}, {Steidelm{\"u}ller}, {Bastian}, {Biermann}, {de
  Torres}, {Gerlach}, {Geyer}, {Hilger}, {Hobbs}, {Lammers}, {McMillan},
  {Stephenson}, {Casta{\~n}eda}, {Davidson}, {Fabricius}, {Gracia-Abril},
  {Portell}, {Rowell}, {Teyssier}, {Torra}, {Bartolom{\'e}}, {Clotet},
  {Garralda}, {Gonz{\'a}lez-Vidal}, {Torra}, {Abbas}, {Altmann}, {Anglada
  Varela}, {Balaguer-N{\'u}{\~n}ez}, {Balog}, {Barache}, {Becciani}, {Bernet},
  {Bertone}, {Bianchi}, {Bouquillon}, {Brown}, {Bucciarelli}, {Busonero},
  {Butkevich}, {Buzzi}, {Cancelliere}, {Carlucci}, {Charlot}, {Cioni},
  {Crosta}, {Crowley}, {del Peloso}, {del Pozo}, {Drimmel}, {Esquej}, {Fienga},
  {Fraile}, {Gai}, {Garcia-Reinaldos}, {Guerra}, {Hambly}, {Hauser},
  {Jan{\ss}en}, {Jordan}, {Kostrzewa-Rutkowska}, {Lattanzi}, {Liao}, {Licata},
  {Lister}, {L{\"o}ffler}, {Marchant}, {Masip}, {Mignard}, {Mints}, {Molina},
  {Mora}, {Morbidelli}, {Murphy}, {Pagani}, {Panuzzo}, {Pe{\~n}alosa Esteller},
  {Poggio}, {Re Fiorentin}, {Riva}, {Sagrist{\`a} Sell{\'e}s}, {Sanchez
  Gimenez}, {Sarasso}, {Sciacca}, {Siddiqui}, {Smart}, {Souami}, {Spagna},
  {Steele}, {Taris}, {Utrilla}, {van Reeven}, \& {Vecchiato}}]{Lindegren2021}
{Lindegren}, L., {Klioner}, S.~A., {Hern{\'a}ndez}, J., {et~al.} 2021, \aap,
  649, A2

\bibitem[{{Mahadevan} {et~al.}(2012){Mahadevan}, {Ramsey}, {Bender}, {Terrien},
  {Wright}, {Halverson}, {Hearty}, {Nelson}, {Burton}, {Redman}, {Osterman},
  {Diddams}, {Kasting}, {Endl}, \& {Deshpande}}]{2012SPIE.8446E..1SM}
{Mahadevan}, S., {Ramsey}, L., {Bender}, C., {et~al.} 2012, in Society of
  Photo-Optical Instrumentation Engineers (SPIE) Conference Series, Vol. 8446,
  Ground-based and Airborne Instrumentation for Astronomy IV, ed. I.~S.
  {McLean}, S.~K. {Ramsay}, \& H.~{Takami}, 84461S

\bibitem[{{Mahadevan} {et~al.}(2021){Mahadevan}, {Stef{\'a}nsson}, {Robertson},
  {Terrien}, {Ninan}, {Holcomb}, {Halverson}, {Cochran}, {Kanodia}, {Ramsey},
  {Wolszczan}, {Endl}, {Bender}, {Diddams}, {Fredrick}, {Hearty}, {Monson},
  {Metcalf}, {Roy}, \& {Schwab}}]{2021ApJ...919L...9M}
{Mahadevan}, S., {Stef{\'a}nsson}, G., {Robertson}, P., {et~al.} 2021, \apjl,
  919, L9

\bibitem[{{Marfil} {et~al.}(2021){Marfil}, {Tabernero}, {Montes}, {Caballero},
  {L{\'a}zaro}, {Gonz{\'a}lez Hern{\'a}ndez}, {Nagel}, {Passegger},
  {Schweitzer}, {Ribas}, {Reiners}, {Quirrenbach}, {Amado}, {Cifuentes},
  {Cort{\'e}s-Contreras}, {Dreizler}, {Duque-Arribas},
  {Galad{\'\i}-Enr{\'\i}quez}, {Henning}, {Jeffers}, {Kaminski}, {K{\"u}rster},
  {Lafarga}, {L{\'o}pez-Gallifa}, {Morales}, {Shan}, \&
  {Zechmeister}}]{2021A&A...656A.162M}
{Marfil}, E., {Tabernero}, H.~M., {Montes}, D., {et~al.} 2021, \aap, 656, A162

\bibitem[{{Mayor} {et~al.}(2003){Mayor}, {Pepe}, {Queloz}, {Bouchy},
  {Rupprecht}, {Lo Curto}, {Avila}, {Benz}, {Bertaux}, {Bonfils}, {Dall},
  {Dekker}, {Delabre}, {Eckert}, {Fleury}, {Gilliotte}, {Gojak}, {Guzman},
  {Kohler}, {Lizon}, {Longinotti}, {Lovis}, {Megevand}, {Pasquini}, {Reyes},
  {Sivan}, {Sosnowska}, {Soto}, {Udry}, {van Kesteren}, {Weber}, \&
  {Weilenmann}}]{2003Msngr.114...20M}
{Mayor}, M., {Pepe}, F., {Queloz}, D., {et~al.} 2003, The Messenger, 114, 20

\bibitem[{{Mayor} \& {Queloz}(1995)}]{1995Natur.378..355M}
{Mayor}, M. \& {Queloz}, D. 1995, \nat, 378, 355

\bibitem[{{Nagel} {et~al.}(2022){Nagel}, {Czesla}, {Kaminski}, {Zechmeister},
  \& {Tal-Or}}]{2022nagel}
{Nagel}, E., {Czesla}, S., {Kaminski}, A., {Zechmeister}, M., \& {Tal-Or}, L.
  2022, \aap, submitted

\bibitem[{{Newton} {et~al.}(2016){Newton}, {Irwin}, {Charbonneau},
  {Berta-Thompson}, {Dittmann}, \& {West}}]{2016ApJ...821...93N}
{Newton}, E.~R., {Irwin}, J., {Charbonneau}, D., {et~al.} 2016, \apj, 821, 93

\bibitem[{{Penoyre} {et~al.}(2020){Penoyre}, {Belokurov}, {Wyn Evans},
  {Everall}, \& {Koposov}}]{Penoyre2020}
{Penoyre}, Z., {Belokurov}, V., {Wyn Evans}, N., {Everall}, A., \& {Koposov},
  S.~E. 2020, \mnras, 495, 321

\bibitem[{{Pepe} {et~al.}(2010){Pepe}, {Cristiani}, {Rebolo Lopez}, {Santos},
  {Amorim}, {Avila}, {Benz}, {Bonifacio}, {Cabral}, {Carvas}, {Cirami},
  {Coelho}, {Comari}, {Coretti}, {De Caprio}, {Dekker}, {Delabre}, {Di
  Marcantonio}, {D'Odorico}, {Fleury}, {Garc{\'\i}a}, {Herreros Linares},
  {Hughes}, {Iwert}, {Lima}, {Lizon}, {Lo Curto}, {Lovis}, {Manescau},
  {Martins}, {M{\'e}gevand}, {Moitinho}, {Molaro}, {Monteiro}, {Monteiro},
  {Pasquini}, {Mordasini}, {Queloz}, {Rasilla}, {Rebord{\~a}o}, {Santana
  Tschudi}, {Santin}, {Sosnowska}, {Span{\`o}}, {Tenegi}, {Udry}, {Vanzella},
  {Viel}, {Zapatero Osorio}, \& {Zerbi}}]{2010SPIE.7735E..0FP}
{Pepe}, F.~A., {Cristiani}, S., {Rebolo Lopez}, R., {et~al.} 2010, in Society
  of Photo-Optical Instrumentation Engineers (SPIE) Conference Series, Vol.
  7735, Ground-based and Airborne Instrumentation for Astronomy III, ed. I.~S.
  {McLean}, S.~K. {Ramsay}, \& H.~{Takami}, 77350F

\bibitem[{{Perdelwitz} {et~al.}(2021){Perdelwitz}, {Mittag}, {Tal-Or},
  {Schmitt}, {Caballero}, {Jeffers}, {Reiners}, {Schweitzer}, {Trifonov},
  {Ribas}, {Quirrenbach}, {Amado}, {Seifert}, {Cifuentes},
  {Cort{\'e}s-Contreras}, {Montes}, {Revilla}, \&
  {Skrzypinski}}]{2021A&A...652A.116P}
{Perdelwitz}, V., {Mittag}, M., {Tal-Or}, L., {et~al.} 2021, \aap, 652, A116

\bibitem[{{P{\'e}rez-Torres} {et~al.}(2021){P{\'e}rez-Torres}, {G{\'o}mez},
  {Ortiz}, {Leto}, {Anglada}, {G{\'o}mez}, {Rodr{\'\i}guez}, {Trigilio},
  {Amado}, {Alberdi}, {Anglada-Escud{\'e}}, {Osorio}, {Umana}, {Berdi{\~n}as},
  {L{\'o}pez-Gonz{\'a}lez}, {Morales}, {Rodr{\'\i}guez-L{\'o}pez}, \&
  {Chibueze}}]{2021A&A...645A..77P}
{P{\'e}rez-Torres}, M., {G{\'o}mez}, J.~F., {Ortiz}, J.~L., {et~al.} 2021,
  \aap, 645, A77

\bibitem[{{Perger} {et~al.}(2021{\natexlab{a}}){Perger}, {Anglada-Escud{\'e}},
  {Ribas}, {Rosich}, {Herrero}, \& {Morales}}]{2021A&A...645A..58P}
{Perger}, M., {Anglada-Escud{\'e}}, G., {Ribas}, I., {et~al.}
  2021{\natexlab{a}}, \aap, 645, A58

\bibitem[{{Perger} {et~al.}(2021{\natexlab{b}}){Perger}, {Ribas},
  {Anglada-Escud{\'e}}, {Morales}, {Amado}, {Caballero}, {Quirrenbach},
  {Reiners}, {B{\'e}jar}, {Dreizler}, {Galad{\'\i}-Enr{\'\i}quez}, {Hatzes},
  {Henning}, {Jeffers}, {Kaminski}, {K{\"u}rster}, {Lafarga}, {Montes},
  {Pall{\'e}}, {Rodr{\'\i}guez-L{\'o}pez}, {Schweitzer}, {Zapatero Osorio}, \&
  {Zechmeister}}]{2021A&A...649L..12P}
{Perger}, M., {Ribas}, I., {Anglada-Escud{\'e}}, G., {et~al.}
  2021{\natexlab{b}}, \aap, 649, L12

\bibitem[{{Perryman}(2018)}]{2018exha.book.....P}
{Perryman}, M. 2018, {The Exoplanet Handbook}

\bibitem[{{Pope} {et~al.}(2020){Pope}, {Bedell}, {Callingham}, {Vedantham},
  {Snellen}, {Price-Whelan}, \& {Shimwell}}]{2020ApJ...890L..19P}
{Pope}, B. J.~S., {Bedell}, M., {Callingham}, J.~R., {et~al.} 2020, \apjl, 890,
  L19

\bibitem[{{Pope} {et~al.}(2021){Pope}, {Callingham}, {Feinstein},
  {G{\"u}nther}, {Vedantham}, {Ansdell}, \& {Shimwell}}]{2021ApJ...919L..10P}
{Pope}, B. J.~S., {Callingham}, J.~R., {Feinstein}, A.~D., {et~al.} 2021,
  \apjl, 919, L10

\bibitem[{{Quirrenbach} {et~al.}(2014){Quirrenbach}, {Amado}, {Caballero},
  {Mundt}, {Reiners}, {Ribas}, {Seifert}, {Abril}, {Aceituno},
  {Alonso-Floriano}, {Ammler-von Eiff}, {Antona Jim{\'e}nez},
  {Anwand-Heerwart}, {Azzaro}, {Bauer}, {Barrado}, {Becerril}, {B{\'e}jar},
  {Ben{\'\i}tez}, {Berdi{\~n}as}, {C{\'a}rdenas}, {Casal}, {Claret},
  {Colom{\'e}}, {Cort{\'e}s-Contreras}, {Czesla}, {Doellinger}, {Dreizler},
  {Feiz}, {Fern{\'a}ndez}, {Galad{\'\i}}, {G{\'a}lvez-Ortiz},
  {Garc{\'\i}a-Piquer}, {Garc{\'\i}a-Vargas}, {Garrido}, {Gesa}, {G{\'o}mez
  Galera}, {Gonz{\'a}lez {\'A}lvarez}, {Gonz{\'a}lez Hern{\'a}ndez},
  {Gr{\"o}zinger}, {Gu{\`a}rdia}, {Guenther}, {de Guindos},
  {Guti{\'e}rrez-Soto}, {Hagen}, {Hatzes}, {Hauschildt}, {Helmling}, {Henning},
  {Hermann}, {Hern{\'a}ndez Casta{\~n}o}, {Herrero}, {Hidalgo}, {Holgado},
  {Huber}, {Huber}, {Jeffers}, {Joergens}, {de Juan}, {Kehr}, {Klein},
  {K{\"u}rster}, {Lamert}, {Lalitha}, {Laun}, {Lemke}, {Lenzen}, {L{\'o}pez del
  Fresno}, {L{\'o}pez Mart{\'\i}}, {L{\'o}pez-Santiago}, {Mall}, {Mandel},
  {Mart{\'\i}n}, {Mart{\'\i}n-Ruiz}, {Mart{\'\i}nez-Rodr{\'\i}guez}, {Marvin},
  {Mathar}, {Mirabet}, {Montes}, {Morales Mu{\~n}oz}, {Moya}, {Naranjo},
  {Ofir}, {Oreiro}, {Pall{\'e}}, {Panduro}, {Passegger}, {P{\'e}rez-Calpena},
  {P{\'e}rez Medialdea}, {Perger}, {Pluto}, {Ram{\'o}n}, {Rebolo}, {Redondo},
  {Reffert}, {Reinhardt}, {Rhode}, {Rix}, {Rodler}, {Rodr{\'\i}guez},
  {Rodr{\'\i}guez-L{\'o}pez}, {Rodr{\'\i}guez-P{\'e}rez}, {Rohloff}, {Rosich},
  {S{\'a}nchez-Blanco}, {S{\'a}nchez Carrasco}, {Sanz-Forcada}, {Sarmiento},
  {Sch{\"a}fer}, {Schiller}, {Schmidt}, {Schmitt}, {Solano}, {Stahl}, {Storz},
  {St{\"u}rmer}, {Su{\'a}rez}, {Ulbrich}, {Veredas}, {Wagner}, {Winkler},
  {Zapatero Osorio}, {Zechmeister}, {Abell{\'a}n de Paco},
  {Anglada-Escud{\'e}}, {del Burgo}, {Klutsch}, {Lizon}, {L{\'o}pez-Morales},
  {Morales}, {Perryman}, {Tulloch}, \& {Xu}}]{2014SPIE.9147E..1FQ}
{Quirrenbach}, A., {Amado}, P.~J., {Caballero}, J.~A., {et~al.} 2014, in
  Society of Photo-Optical Instrumentation Engineers (SPIE) Conference Series,
  Vol. 9147, Ground-based and Airborne Instrumentation for Astronomy V, ed.
  S.~K. {Ramsay}, I.~S. {McLean}, \& H.~{Takami}, 91471F

\bibitem[{{Quirrenbach} {et~al.}(2020){Quirrenbach}, {CARMENES Consortium},
  {Amado}, {Ribas}, {Reiners}, {Caballero}, {Aceituno}, {Alacid},
  {Alonso-Floriano}, {Anglada-Escud{\'e}}, {Azzaro}, {Baroch}, {Bauer},
  {Becerril}, {B{\'e}jar}, {Bluhm}, {Calvo Ortega}, {Cardona Guill{\'e}n},
  {Casasayas-Barris}, {Chaturvedi}, {Cifuentes}, {Colom{\'e}}, {Conte},
  {Cort{\'e}s-Contreras}, {Czesla}, {D{\'\i}ez-Alonso}, {Dom{\'\i}nguez
  Fern{\'a}ndez}, {Dreizler}, {Duque-Arribas}, {Espinoza}, {Fuhrmeister},
  {Galad{\'\i}-Enr{\'\i}quez}, {Gar{\textasciiacute}a Quintana},
  {Gonz{\'a}lez-Alvare}, {Gonz{\'a}lez Cuesta}, {Gonz{\'a}lez Hern{\'a}ndez},
  {Guenther}, {de Guindos}, {Hatzes}, {Henning}, {Herbort}, {Herrero}, {Hintz},
  {Iglesias-P{\'a}ra}, {Jeffers}, {Johnson}, {de Juan}, {Kaminski}, {Kemmer},
  {Khaimova}, {Khalafinejad}, {Klahr}, {Kossakowski}, {Kreidberg},
  {K{\"u}rster}, {Labarga}, {Lafarga}, {Lamp{\'o}n}, {Lara}, {Lillo-Box},
  {Lodieu}, {L{\'o}pez Gallifa}, {L{\'o}pez Gonz{\'a}lez}, {L{\'o}pez-Puertas},
  {Luque}, {Marfil}, {Mart{\'\i}n-Ruiz}, {Matth{\'e}}, {Molaverdikhani},
  {Montes}, {Morales}, {Morales-Calder{\'o}on}, {Nagel}, {Nortmann}, {Nowak},
  {Ofir}, {Oshaghi}, {Pall{\'e}}, {Passegger}, {Pavlov}, {Pedraz},
  {Perdelwitz}, {Perger}, {Reffert}, {Revilla}, {Rodr{\'\i}guez},
  {Rodr{\'\i}guez L{\'o}pez}, {Sabotta}, {Sadegi}, {Sairam}, {Salz},
  {S{\'a}nchez-L{\'o}pez}, {Sanz-Forcada}, {Sarkis}, {Sch{\"a}fer}, {Schiller},
  {Schlecker}, {Schmitt}, {Sch{\"o}fer}, {Schweitzer}, {Seiferta}, {Shan},
  {Shulyak}, {Skrzypinski}, {Solano}, {Soto}, {Stahl}, {Stangret}, {Stock},
  {Strachan}, {Stuber}, {St{\"u}rmer}, {Tabernero}, {Tal-Or}, {Tala-Pinto},
  {Trifonov}, {Vanaverbeke}, {Yan}, {Zapatero Osorio}, \&
  {Zechmeister}}]{2020SPIE11447E..3CQ}
{Quirrenbach}, A., {CARMENES Consortium}, {Amado}, P.~J., {et~al.} 2020, in
  Society of Photo-Optical Instrumentation Engineers (SPIE) Conference Series,
  11447, 114473C

\bibitem[{{Rajpaul} {et~al.}(2021){Rajpaul}, {Buchhave}, {Lacedelli}, {Rice},
  {Mortier}, {Malavolta}, {Aigrain}, {Borsato}, {Mayo}, {Charbonneau},
  {Damasso}, {Dumusque}, {Ghedina}, {Latham}, {L{\'o}pez-Morales},
  {Magazz{\`u}}, {Micela}, {Molinari}, {Pepe}, {Piotto}, {Poretti}, {Rowther},
  {Sozzetti}, {Udry}, \& {Watson}}]{2021MNRAS.507.1847R}
{Rajpaul}, V.~M., {Buchhave}, L.~A., {Lacedelli}, G., {et~al.} 2021, \mnras,
  507, 1847

\bibitem[{{Reiners} {et~al.}(2018{\natexlab{a}}){Reiners}, {Ribas},
  {Zechmeister}, {Caballero}, {Trifonov}, {Dreizler}, {Morales}, {Tal-Or},
  {Lafarga}, {Quirrenbach}, {Amado}, {Kaminski}, {Jeffers}, {Aceituno},
  {B{\'e}jar}, {Gu{\`a}rdia}, {Guenther}, {Hagen}, {Montes}, {Passegger},
  {Seifert}, {Schweitzer}, {Cort{\'e}s-Contreras}, {Abril}, {Alonso-Floriano},
  {Ammler-von Eiff}, {Antona}, {Anglada-Escud{\'e}}, {Anwand-Heerwart},
  {Arroyo-Torres}, {Azzaro}, {Baroch}, {Barrado}, {Bauer}, {Becerril},
  {Ben{\'\i}tez}, {Berdi{\~n}as}, {Bergond}, {Bl{\"u}mcke}, {Brinkm{\"o}ller},
  {del Burgo}, {Cano}, {C{\'a}rdenas V{\'a}zquez}, {Casal}, {Cifuentes},
  {Claret}, {Colom{\'e}}, {Czesla}, {D{\'\i}ez-Alonso}, {Feiz},
  {Fern{\'a}ndez}, {Ferro}, {Fuhrmeister}, {Galad{\'\i}-Enr{\'\i}quez},
  {Garcia-Piquer}, {Garc{\'\i}a Vargas}, {Gesa}, {G{\'o}mez Galera},
  {Gonz{\'a}lez Hern{\'a}ndez}, {Gonz{\'a}lez-Peinado}, {Gr{\"o}zinger},
  {Grohnert}, {Guijarro}, {de Guindos}, {Guti{\'e}rrez-Soto}, {Hatzes},
  {Hauschildt}, {Hedrosa}, {Helmling}, {Henning}, {Hermelo}, {Hern{\'a}ndez
  Arab{\'\i}}, {Hern{\'a}ndez Casta{\~n}o}, {Hern{\'a}ndez Hernando},
  {Herrero}, {Huber}, {Huke}, {Johnson}, {de Juan}, {Kim}, {Klein},
  {Kl{\"u}ter}, {Klutsch}, {K{\"u}rster}, {Labarga}, {Lamert}, {Lamp{\'o}n},
  {Lara}, {Laun}, {Lemke}, {Lenzen}, {Launhardt}, {L{\'o}pez del Fresno},
  {L{\'o}pez-Gonz{\'a}lez}, {L{\'o}pez-Puertas}, {L{\'o}pez Salas},
  {L{\'o}pez-Santiago}, {Luque}, {Mag{\'a}n Madinabeitia}, {Mall}, {Mancini},
  {Mandel}, {Marfil}, {Mar{\'\i}n Molina}, {Maroto Fern{\'a}ndez},
  {Mart{\'\i}n}, {Mart{\'\i}n-Ruiz}, {Marvin}, {Mathar}, {Mirabet},
  {Moreno-Raya}, {Moya}, {Mundt}, {Nagel}, {Naranjo}, {Nortmann}, {Nowak},
  {Ofir}, {Oreiro}, {Pall{\'e}}, {Panduro}, {Pascual}, {Pavlov}, {Pedraz},
  {P{\'e}rez-Calpena}, {P{\'e}rez Medialdea}, {Perger}, {Perryman}, {Pluto},
  {Rabaza}, {Ram{\'o}n}, {Rebolo}, {Redondo}, {Reffert}, {Reinhart}, {Rhode},
  {Rix}, {Rodler}, {Rodr{\'\i}guez}, {Rodr{\'\i}guez-L{\'o}pez},
  {Rodr{\'\i}guez Trinidad}, {Rohloff}, {Rosich}, {Sadegi},
  {S{\'a}nchez-Blanco}, {S{\'a}nchez Carrasco}, {S{\'a}nchez-L{\'o}pez},
  {Sanz-Forcada}, {Sarkis}, {Sarmiento}, {Sch{\"a}fer}, {Schmitt}, {Schiller},
  {Sch{\"o}fer}, {Solano}, {Stahl}, {Strachan}, {St{\"u}rmer}, {Su{\'a}rez},
  {Tabernero}, {Tala}, {Tulloch}, {Ulbrich}, {Veredas}, {Vico Linares},
  {Vilardell}, {Wagner}, {Winkler}, {Wolthoff}, {Xu}, {Yan}, \& {Zapatero
  Osorio}}]{2018A&A...609L...5R}
{Reiners}, A., {Ribas}, I., {Zechmeister}, M., {et~al.} 2018{\natexlab{a}},
  \aap, 609, L5

\bibitem[{{Reiners} {et~al.}(2022){Reiners}, {Shulyak}, {K{\"a}pyl{\"a}},
  {Ribas}, {Nagel}, {Zechmeister}, {Caballero}, {Shan}, {Fuhrmeister},
  {Quirrenbach}, {Amado}, {Montes}, {Jeffers}, {Azzaro}, {B{\'e}jar},
  {Chaturvedi}, {Henning}, {K{\"u}rster}, \& {Pall{\'e}}}]{2022A&A...662A..41R}
{Reiners}, A., {Shulyak}, D., {K{\"a}pyl{\"a}}, P.~J., {et~al.} 2022, \aap,
  662, A41

\bibitem[{{Reiners} {et~al.}(2018{\natexlab{b}}){Reiners}, {Zechmeister},
  {Caballero}, {Ribas}, {Morales}, {Jeffers}, {Sch{\"o}fer}, {Tal-Or},
  {Quirrenbach}, {Amado}, {Kaminski}, {Seifert}, {Abril}, {Aceituno},
  {Alonso-Floriano}, {Ammler-von Eiff}, {Antona}, {Anglada-Escud{\'e}},
  {Anwand-Heerwart}, {Arroyo-Torres}, {Azzaro}, {Baroch}, {Barrado}, {Bauer},
  {Becerril}, {B{\'e}jar}, {Ben{\'\i}tez}, {Berdinas}, {Bergond},
  {Bl{\"u}mcke}, {Brinkm{\"o}ller}, {del Burgo}, {Cano}, {C{\'a}rdenas
  V{\'a}zquez}, {Casal}, {Cifuentes}, {Claret}, {Colom{\'e}},
  {Cort{\'e}s-Contreras}, {Czesla}, {D{\'\i}ez-Alonso}, {Dreizler}, {Feiz},
  {Fern{\'a}ndez}, {Ferro}, {Fuhrmeister}, {Galad{\'\i}-Enr{\'\i}quez},
  {Garcia-Piquer}, {Garc{\'\i}a Vargas}, {Gesa}, {G{\'o}mez Galera},
  {Gonz{\'a}lez Hern{\'a}ndez}, {Gonz{\'a}lez-Peinado}, {Gr{\"o}zinger},
  {Grohnert}, {Gu{\`a}rdia}, {Guenther}, {Guijarro}, {de Guindos},
  {Guti{\'e}rrez-Soto}, {Hagen}, {Hatzes}, {Hauschildt}, {Hedrosa}, {Helmling},
  {Henning}, {Hermelo}, {Hern{\'a}ndez Arab{\'\i}}, {Hern{\'a}ndez
  Casta{\~n}o}, {Hern{\'a}ndez Hernando}, {Herrero}, {Huber}, {Huke},
  {Johnson}, {de Juan}, {Kim}, {Klein}, {Kl{\"u}ter}, {Klutsch}, {K{\"u}rster},
  {Lafarga}, {Lamert}, {Lamp{\'o}n}, {Lara}, {Laun}, {Lemke}, {Lenzen},
  {Launhardt}, {L{\'o}pez del Fresno}, {L{\'o}pez-Gonz{\'a}lez},
  {L{\'o}pez-Puertas}, {L{\'o}pez Salas}, {L{\'o}pez-Santiago}, {Luque},
  {Mag{\'a}n Madinabeitia}, {Mall}, {Mancini}, {Mandel}, {Marfil}, {Mar{\'\i}n
  Molina}, {Maroto Fern{\'a}ndez}, {Mart{\'\i}n}, {Mart{\'\i}n-Ruiz}, {Marvin},
  {Mathar}, {Mirabet}, {Montes}, {Moreno-Raya}, {Moya}, {Mundt}, {Nagel},
  {Naranjo}, {Nortmann}, {Nowak}, {Ofir}, {Oreiro}, {Pall{\'e}}, {Panduro},
  {Pascual}, {Passegger}, {Pavlov}, {Pedraz}, {P{\'e}rez-Calpena}, {P{\'e}rez
  Medialdea}, {Perger}, {Perryman}, {Pluto}, {Rabaza}, {Ram{\'o}n}, {Rebolo},
  {Redondo}, {Reffert}, {Reinhart}, {Rhode}, {Rix}, {Rodler}, {Rodr{\'\i}guez},
  {Rodr{\'\i}guez-L{\'o}pez}, {Rodr{\'\i}guez Trinidad}, {Rohloff}, {Rosich},
  {Sadegi}, {S{\'a}nchez-Blanco}, {S{\'a}nchez Carrasco},
  {S{\'a}nchez-L{\'o}pez}, {Sanz-Forcada}, {Sarkis}, {Sarmiento},
  {Sch{\"a}fer}, {Schmitt}, {Schiller}, {Schweitzer}, {Solano}, {Stahl},
  {Strachan}, {St{\"u}rmer}, {Su{\'a}rez}, {Tabernero}, {Tala}, {Trifonov},
  {Tulloch}, {Ulbrich}, {Veredas}, {Vico Linares}, {Vilardell}, {Wagner},
  {Winkler}, {Wolthoff}, {Xu}, {Yan}, \& {Zapatero
  Osorio}}]{2018A&A...612A..49R}
{Reiners}, A., {Zechmeister}, M., {Caballero}, J.~A., {et~al.}
  2018{\natexlab{b}}, \aap, 612, A49

\bibitem[{{Ribas} {et~al.}(2022){Ribas}, {Reiners}, {Zechmeister}, {Caballero},
  {Morales}, {Morales}, {Morales}, {Morales}, {Morales}, \&
  {Morales}}]{2022ribas}
{Ribas}, I., {Reiners}, A., {Zechmeister}, M., {et~al.} 2022, \aap, submitted

\bibitem[{{Ricker} {et~al.}(2015){Ricker}, {Winn}, {Vanderspek}, {Latham},
  {Bakos}, {Bean}, {Berta-Thompson}, {Brown}, {Buchhave}, {Butler}, {Butler},
  {Chaplin}, {Charbonneau}, {Christensen-Dalsgaard}, {Clampin}, {Deming},
  {Doty}, {De Lee}, {Dressing}, {Dunham}, {Endl}, {Fressin}, {Ge}, {Henning},
  {Holman}, {Howard}, {Ida}, {Jenkins}, {Jernigan}, {Johnson}, {Kaltenegger},
  {Kawai}, {Kjeldsen}, {Laughlin}, {Levine}, {Lin}, {Lissauer}, {MacQueen},
  {Marcy}, {McCullough}, {Morton}, {Narita}, {Paegert}, {Palle}, {Pepe},
  {Pepper}, {Quirrenbach}, {Rinehart}, {Sasselov}, {Sato}, {Seager},
  {Sozzetti}, {Stassun}, {Sullivan}, {Szentgyorgyi}, {Torres}, {Udry}, \&
  {Villasenor}}]{2015JATIS...1a4003R}
{Ricker}, G.~R., {Winn}, J.~N., {Vanderspek}, R., {et~al.} 2015, Journal of
  Astronomical Telescopes, Instruments, and Systems, 1, 014003

\bibitem[{{Saur} {et~al.}(2013){Saur}, {Grambusch}, {Duling}, {Neubauer}, \&
  {Simon}}]{saur13}
{Saur}, J., {Grambusch}, T., {Duling}, S., {Neubauer}, F.~M., \& {Simon}, S.
  2013, \aap, 552, A119

\bibitem[{{Sch{\"o}fer} {et~al.}(2019){Sch{\"o}fer}, {Jeffers}, {Reiners},
  {Shulyak}, {Fuhrmeister}, {Johnson}, {Zechmeister}, {Ribas}, {Quirrenbach},
  {Amado}, {Caballero}, {Anglada-Escud{\'e}}, {Bauer}, {B{\'e}jar},
  {Cort{\'e}s-Contreras}, {Dreizler}, {Guenther}, {Kaminski}, {K{\"u}rster},
  {Lafarga}, {Montes}, {Morales}, {Pedraz}, \& {Tal-Or}}]{2019A&A...623A..44S}
{Sch{\"o}fer}, P., {Jeffers}, S.~V., {Reiners}, A., {et~al.} 2019, \aap, 623,
  A44

\bibitem[{{Schweitzer} {et~al.}(2019){Schweitzer}, {Passegger}, {Cifuentes},
  {B{\'e}jar}, {Cort{\'e}s-Contreras}, {Caballero}, {del Burgo}, {Czesla},
  {K{\"u}rster}, {Montes}, {Zapatero Osorio}, {Ribas}, {Reiners},
  {Quirrenbach}, {Amado}, {Aceituno}, {Anglada-Escud{\'e}}, {Bauer},
  {Dreizler}, {Jeffers}, {Guenther}, {Henning}, {Kaminski}, {Lafarga},
  {Marfil}, {Morales}, {Schmitt}, {Seifert}, {Solano}, {Tabernero}, \&
  {Zechmeister}}]{2019A&A...625A..68S}
{Schweitzer}, A., {Passegger}, V.~M., {Cifuentes}, C., {et~al.} 2019, \aap,
  625, A68

\bibitem[{{Shimwell} {et~al.}(2019){Shimwell}, {Tasse}, {Hardcastle}, {Mechev},
  {Williams}, {Best}, {R{\"o}ttgering}, {Callingham}, {Dijkema}, {de Gasperin},
  {Hoang}, {Hugo}, {Mirmont}, {Oonk}, {Prandoni}, {Rafferty}, {Sabater},
  {Smirnov}, {van Weeren}, {White}, {Atemkeng}, {Bester}, {Bonnassieux},
  {Br{\"u}ggen}, {Brunetti}, {Chy{\.z}y}, {Cochrane}, {Conway}, {Croston},
  {Danezi}, {Duncan}, {Haverkorn}, {Heald}, {Iacobelli}, {Intema}, {Jackson},
  {Jamrozy}, {Jarvis}, {Lakhoo}, {Mevius}, {Miley}, {Morabito}, {Morganti},
  {Nisbet}, {Orr{\'u}}, {Perkins}, {Pizzo}, {Schrijvers}, {Smith}, {Vermeulen},
  {Wise}, {Alegre}, {Bacon}, {van Bemmel}, {Beswick}, {Bonafede}, {Botteon},
  {Bourke}, {Brienza}, {Calistro Rivera}, {Cassano}, {Clarke}, {Conselice},
  {Dettmar}, {Drabent}, {Dumba}, {Emig}, {En{\ss}lin}, {Ferrari}, {Garrett},
  {G{\'e}nova-Santos}, {Goyal}, {G{\"u}rkan}, {Hale}, {Harwood}, {Heesen},
  {Hoeft}, {Horellou}, {Jackson}, {Kokotanekov}, {Kondapally},
  {Kunert-Bajraszewska}, {Mahatma}, {Mahony}, {Mandal}, {McKean}, {Merloni},
  {Mingo}, {Miskolczi}, {Mooney}, {Nikiel-Wroczy{\'n}ski}, {O'Sullivan},
  {Quinn}, {Reich}, {Roskowi{\'n}ski}, {Rowlinson}, {Savini}, {Saxena},
  {Schwarz}, {Shulevski}, {Sridhar}, {Stacey}, {Urquhart}, {van der Wiel},
  {Varenius}, {Webster}, \& {Wilber}}]{2019A&A...622A...1S}
{Shimwell}, T.~W., {Tasse}, C., {Hardcastle}, M.~J., {et~al.} 2019, \aap, 622,
  A1

\bibitem[{{Simkin}(1974)}]{1974A&A....31..129S}
{Simkin}, S.~M. 1974, \aap, 31, 129

\bibitem[{{Sing} {et~al.}(2019){Sing}, {Lavvas}, {Ballester}, {Lecavelier des
  Etangs}, {Marley}, {Nikolov}, {Ben-Jaffel}, {Bourrier}, {Buchhave}, {Deming},
  {Ehrenreich}, {Mikal-Evans}, {Kataria}, {Lewis}, {L{\'o}pez-Morales},
  {Garc{\'\i}a Mu{\~n}oz}, {Henry}, {Sanz-Forcada}, {Spake}, {Wakeford}, \&
  {PanCET Collaboration}}]{2019AJ....158...91S}
{Sing}, D.~K., {Lavvas}, P., {Ballester}, G.~E., {et~al.} 2019, \aj, 158, 91

\bibitem[{{Snellen} {et~al.}(2008){Snellen}, {Albrecht}, {de Mooij}, \& {Le
  Poole}}]{2008A&A...487..357S}
{Snellen}, I.~A.~G., {Albrecht}, S., {de Mooij}, E.~J.~W., \& {Le Poole}, R.~S.
  2008, \aap, 487, 357

\bibitem[{{Stevenson}(2010)}]{2010SSRv..152..651S}
{Stevenson}, D.~J. 2010, \ssr, 152, 651

\bibitem[{{Tinetti} {et~al.}(2007){Tinetti}, {Vidal-Madjar}, {Liang},
  {Beaulieu}, {Yung}, {Carey}, {Barber}, {Tennyson}, {Ribas}, {Allard},
  {Ballester}, {Sing}, \& {Selsis}}]{2007Natur.448..169T}
{Tinetti}, G., {Vidal-Madjar}, A., {Liang}, M.-C., {et~al.} 2007, \nat, 448,
  169

\bibitem[{{Trifonov} {et~al.}(2021){Trifonov}, {Caballero}, {Morales},
  {Seifahrt}, {Ribas}, {Reiners}, {Bean}, {Luque}, {Parviainen}, {Pall{\'e}},
  {Stock}, {Zechmeister}, {Amado}, {Anglada-Escud{\'e}}, {Azzaro}, {Barclay},
  {B{\'e}jar}, {Bluhm}, {Casasayas-Barris}, {Cifuentes}, {Collins}, {Collins},
  {Cort{\'e}s-Contreras}, {de Leon}, {Dreizler}, {Dressing}, {Esparza-Borges},
  {Espinoza}, {Fausnaugh}, {Fukui}, {Hatzes}, {Hellier}, {Henning}, {Henze},
  {Herrero}, {Jeffers}, {Jenkins}, {Jensen}, {Kaminski}, {Kasper},
  {Kossakowski}, {K{\"u}rster}, {Lafarga}, {Latham}, {Mann}, {Molaverdikhani},
  {Montes}, {Montet}, {Murgas}, {Narita}, {Oshagh}, {Passegger}, {Pollacco},
  {Quinn}, {Quirrenbach}, {Ricker}, {Rodr{\'\i}guez L{\'o}pez}, {Sanz-Forcada},
  {Schwarz}, {Schweitzer}, {Seager}, {Shporer}, {Stangret}, {St{\"u}rmer},
  {Tan}, {Tenenbaum}, {Twicken}, {Vanderspek}, \& {Winn}}]{2021Sci...371.1038T}
{Trifonov}, T., {Caballero}, J.~A., {Morales}, J.~C., {et~al.} 2021, Science,
  371, 1038

\bibitem[{{Trotta}(2008)}]{2008ConPh..49...71T}
{Trotta}, R. 2008, Contemporary Physics, 49, 71

\bibitem[{{Turnpenney} {et~al.}(2018){Turnpenney}, {Nichols}, {Wynn}, \&
  {Burleigh}}]{turnpenney18}
{Turnpenney}, S., {Nichols}, J.~D., {Wynn}, G.~A., \& {Burleigh}, M.~R. 2018,
  \apj, 854, 72

\bibitem[{{van Haarlem} {et~al.}(2013){van Haarlem}, {Wise}, {Gunst}, {Heald},
  {McKean}, {Hessels}, {de Bruyn}, {Nijboer}, {Swinbank}, {Fallows},
  {Brentjens}, {Nelles}, {Beck}, {Falcke}, {Fender}, {H{\"o}randel},
  {Koopmans}, {Mann}, {Miley}, {R{\"o}ttgering}, {Stappers}, {Wijers},
  {Zaroubi}, {van den Akker}, {Alexov}, {Anderson}, {Anderson}, {van Ardenne},
  {Arts}, {Asgekar}, {Avruch}, {Batejat}, {B{\"a}hren}, {Bell}, {Bell}, {van
  Bemmel}, {Bennema}, {Bentum}, {Bernardi}, {Best}, {B{\^\i}rzan}, {Bonafede},
  {Boonstra}, {Braun}, {Bregman}, {Breitling}, {van de Brink}, {Broderick},
  {Broekema}, {Brouw}, {Br{\"u}ggen}, {Butcher}, {van Cappellen}, {Ciardi},
  {Coenen}, {Conway}, {Coolen}, {Corstanje}, {Damstra}, {Davies}, {Deller},
  {Dettmar}, {van Diepen}, {Dijkstra}, {Donker}, {Doorduin}, {Dromer}, {Drost},
  {van Duin}, {Eisl{\"o}ffel}, {van Enst}, {Ferrari}, {Frieswijk}, {Gankema},
  {Garrett}, {de Gasperin}, {Gerbers}, {de Geus}, {Grie{\ss}meier}, {Grit},
  {Gruppen}, {Hamaker}, {Hassall}, {Hoeft}, {Holties}, {Horneffer}, {van der
  Horst}, {van Houwelingen}, {Huijgen}, {Iacobelli}, {Intema}, {Jackson},
  {Jelic}, {de Jong}, {Juette}, {Kant}, {Karastergiou}, {Koers}, {Kollen},
  {Kondratiev}, {Kooistra}, {Koopman}, {Koster}, {Kuniyoshi}, {Kramer},
  {Kuper}, {Lambropoulos}, {Law}, {van Leeuwen}, {Lemaitre}, {Loose}, {Maat},
  {Macario}, {Markoff}, {Masters}, {McFadden}, {McKay-Bukowski}, {Meijering},
  {Meulman}, {Mevius}, {Middelberg}, {Millenaar}, {Miller-Jones}, {Mohan},
  {Mol}, {Morawietz}, {Morganti}, {Mulcahy}, {Mulder}, {Munk}, {Nieuwenhuis},
  {van Nieuwpoort}, {Noordam}, {Norden}, {Noutsos}, {Offringa}, {Olofsson},
  {Omar}, {Orr{\'u}}, {Overeem}, {Paas}, {Pandey-Pommier}, {Pandey}, {Pizzo},
  {Polatidis}, {Rafferty}, {Rawlings}, {Reich}, {de Reijer}, {Reitsma},
  {Renting}, {Riemers}, {Rol}, {Romein}, {Roosjen}, {Ruiter}, {Scaife}, {van
  der Schaaf}, {Scheers}, {Schellart}, {Schoenmakers}, {Schoonderbeek},
  {Serylak}, {Shulevski}, {Sluman}, {Smirnov}, {Sobey}, {Spreeuw}, {Steinmetz},
  {Sterks}, {Stiepel}, {Stuurwold}, {Tagger}, {Tang}, {Tasse}, {Thomas},
  {Thoudam}, {Toribio}, {van der Tol}, {Usov}, {van Veelen}, {van der Veen},
  {ter Veen}, {Verbiest}, {Vermeulen}, {Vermaas}, {Vocks}, {Vogt}, {de Vos},
  {van der Wal}, {van Weeren}, {Weggemans}, {Weltevrede}, {White}, {Wijnholds},
  {Wilhelmsson}, {Wucknitz}, {Yatawatta}, {Zarka}, {Zensus}, \& {van
  Zwieten}}]{2013A&A...556A...2V}
{van Haarlem}, M.~P., {Wise}, M.~W., {Gunst}, A.~W., {et~al.} 2013, \aap, 556,
  A2

\bibitem[{{Vedantham} {et~al.}(2020){Vedantham}, {Callingham}, {Shimwell},
  {Tasse}, {Pope}, {Bedell}, {Snellen}, {Best}, {Hardcastle}, {Haverkorn},
  {Mechev}, {O'Sullivan}, {R{\"o}ttgering}, \& {White}}]{2020NatAs...4..577V}
{Vedantham}, H.~K., {Callingham}, J.~R., {Shimwell}, T.~W., {et~al.} 2020,
  Nature Astronomy, 4, 577

\bibitem[{{Vogt} {et~al.}(1994){Vogt}, {Allen}, {Bigelow}, {Bresee}, {Brown},
  {Cantrall}, {Conrad}, {Couture}, {Delaney}, {Epps}, {Hilyard}, {Hilyard},
  {Horn}, {Jern}, {Kanto}, {Keane}, {Kibrick}, {Lewis}, {Osborne},
  {Pardeilhan}, {Pfister}, {Ricketts}, {Robinson}, {Stover}, {Tucker}, {Ward},
  \& {Wei}}]{1994SPIE.2198..362V}
{Vogt}, S.~S., {Allen}, S.~L., {Bigelow}, B.~C., {et~al.} 1994, in Society of
  Photo-Optical Instrumentation Engineers (SPIE) Conference Series, Vol. 2198,
  Instrumentation in Astronomy VIII, ed. D.~L. {Crawford} \& E.~R. {Craine},
  362

\bibitem[{{Yan} {et~al.}(2019){Yan}, {Casasayas-Barris}, {Molaverdikhani},
  {Alonso-Floriano}, {Reiners}, {Pall{\'e}}, {Henning}, {Molli{\`e}re}, {Chen},
  {Nortmann}, {Snellen}, {Ribas}, {Quirrenbach}, {Caballero}, {Amado},
  {Azzaro}, {Bauer}, {Cort{\'e}s Contreras}, {Czesla}, {Khalafinejad}, {Lara},
  {L{\'o}pez-Puertas}, {Montes}, {Nagel}, {Oshagh}, {S{\'a}nchez-L{\'o}pez},
  {Stangret}, \& {Zechmeister}}]{2019A&A...632A..69Y}
{Yan}, F., {Casasayas-Barris}, N., {Molaverdikhani}, K., {et~al.} 2019, \aap,
  632, A69

\bibitem[{{Zarka}(1998)}]{1998JGR...10320159Z}
{Zarka}, P. 1998, \jgr, 103, 20159

\bibitem[{{Zarka}(2007)}]{zarka07}
{Zarka}, P. 2007, \planss, 55, 598

\bibitem[{{Zechmeister} \& {K{\"u}rster}(2009)}]{2009A&A...496..577Z}
{Zechmeister}, M. \& {K{\"u}rster}, M. 2009, \aap, 496, 577

\bibitem[{{Zechmeister} {et~al.}(2009){Zechmeister}, {K{\"u}rster}, \&
  {Endl}}]{2009A&A...505..859Z}
{Zechmeister}, M., {K{\"u}rster}, M., \& {Endl}, M. 2009, \aap, 505, 859

\bibitem[{{Zechmeister} {et~al.}(2018){Zechmeister}, {Reiners}, {Amado},
  {Azzaro}, {Bauer}, {B{\'e}jar}, {Caballero}, {Guenther}, {Hagen}, {Jeffers},
  {Kaminski}, {K{\"u}rster}, {Launhardt}, {Montes}, {Morales}, {Quirrenbach},
  {Reffert}, {Ribas}, {Seifert}, {Tal-Or}, \& {Wolthoff}}]{2018A&A...609A..12Z}
{Zechmeister}, M., {Reiners}, A., {Amado}, P.~J., {et~al.} 2018, \aap, 609, A12

\end{thebibliography}
\begin{appendix}

\section{Additional data plots and periodograms}

\begin{figure*}
\centering
\includegraphics[width=\textwidth]{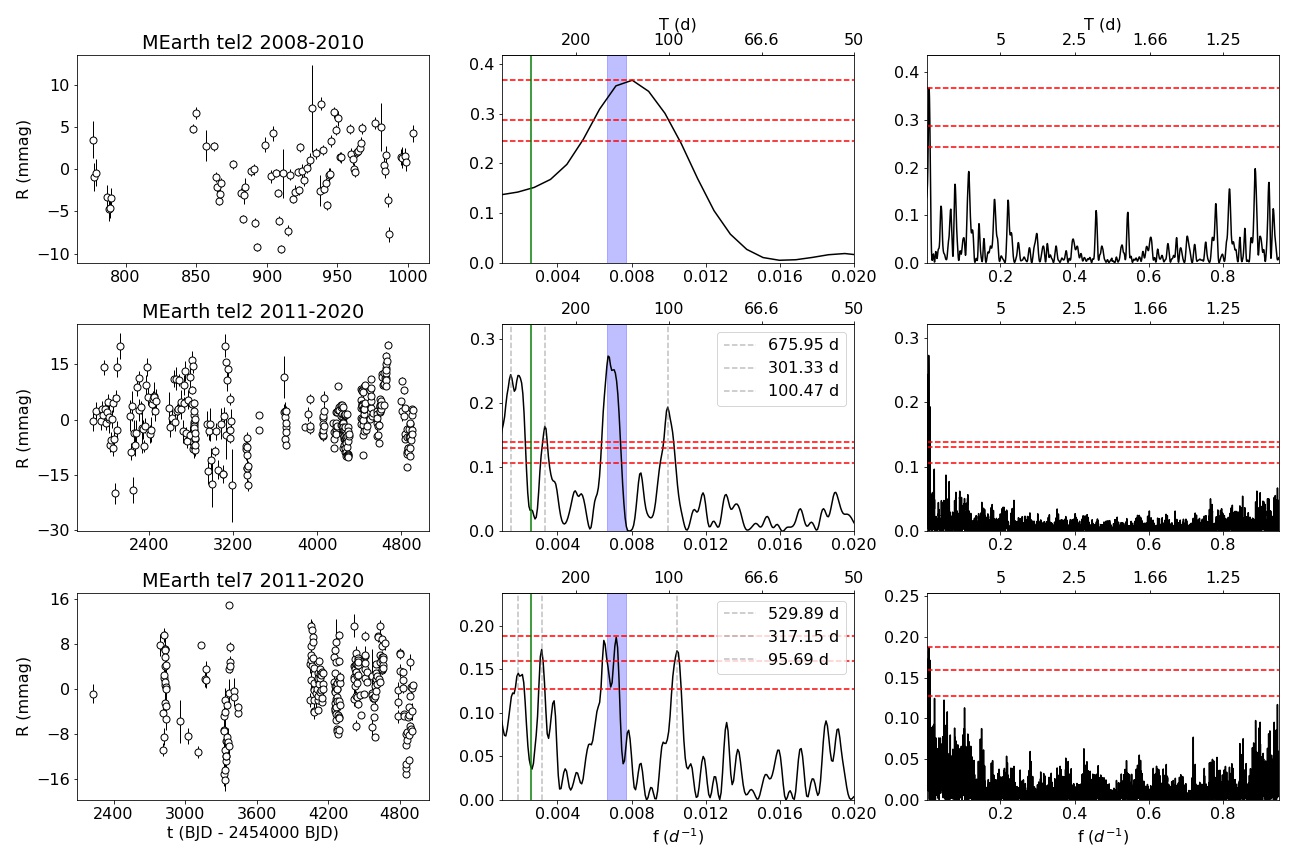} 
\caption{Individual MEarth datasets separated according to observational season and telescope (left panels). We also show  their periodograms for longer periods ($>$50\,days, middle panels) and the full periodograms (right panels). We mark the significant signals in grey, the period of our planet in green, and the rotational period of the star in blue, within the errors. The FAP levels of 10, 1, and 0.1\% are shown by the red dashed lines.} \label{prewhit1}
\end{figure*}

\begin{figure*}
\centering
\includegraphics[width=\textwidth]{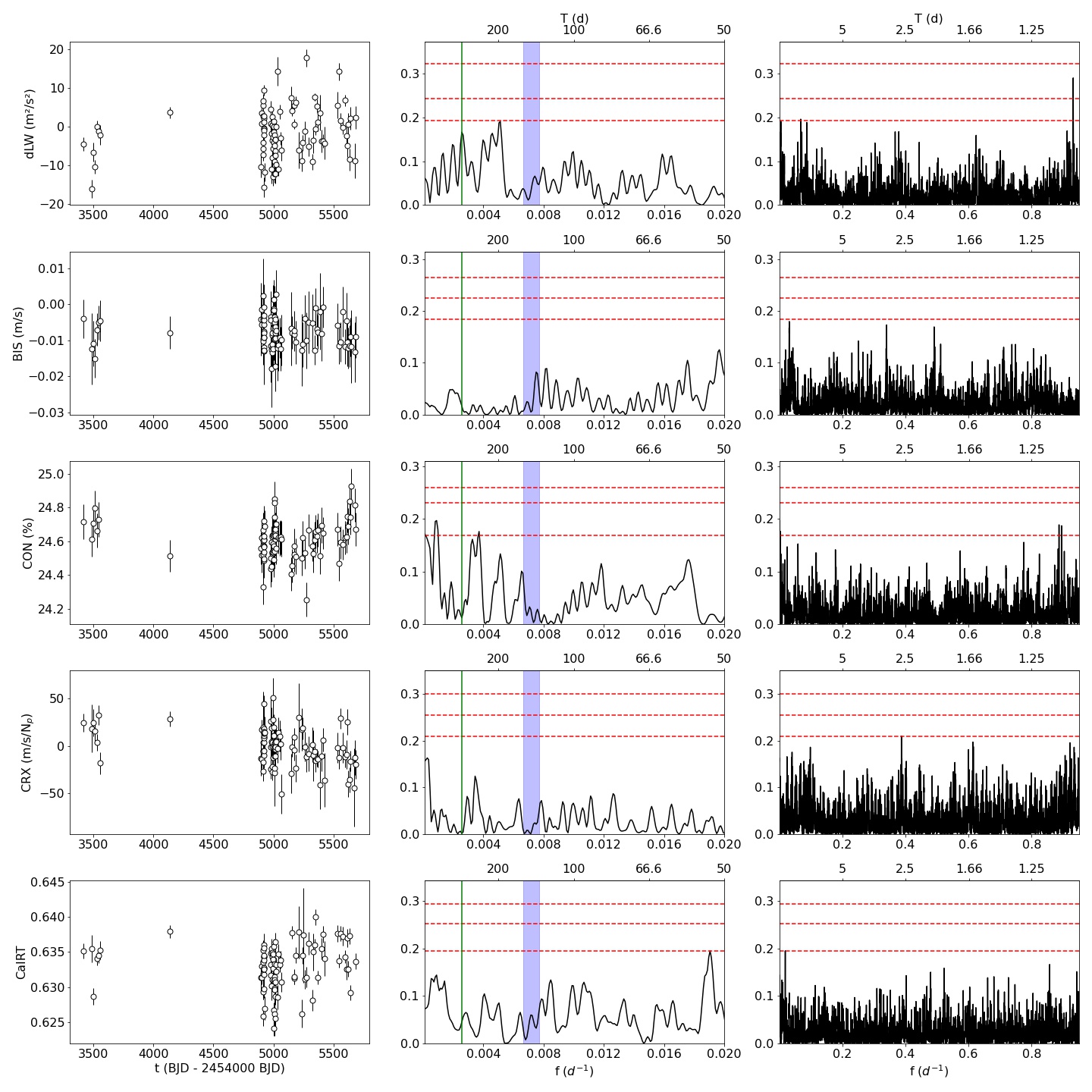}
 \caption{Spectroscopic index time-series data from CARMENES, which do not show any significant signals. From top to bottom this is as indicated dLw, BIS, CON, CRX, and CaIRT indices. Further details are given in Fig.\,\ref{prewhit1}.} \label{prewhit2} 
\end{figure*}

\begin{figure*}
\centering
\includegraphics[width=0.66\textwidth]{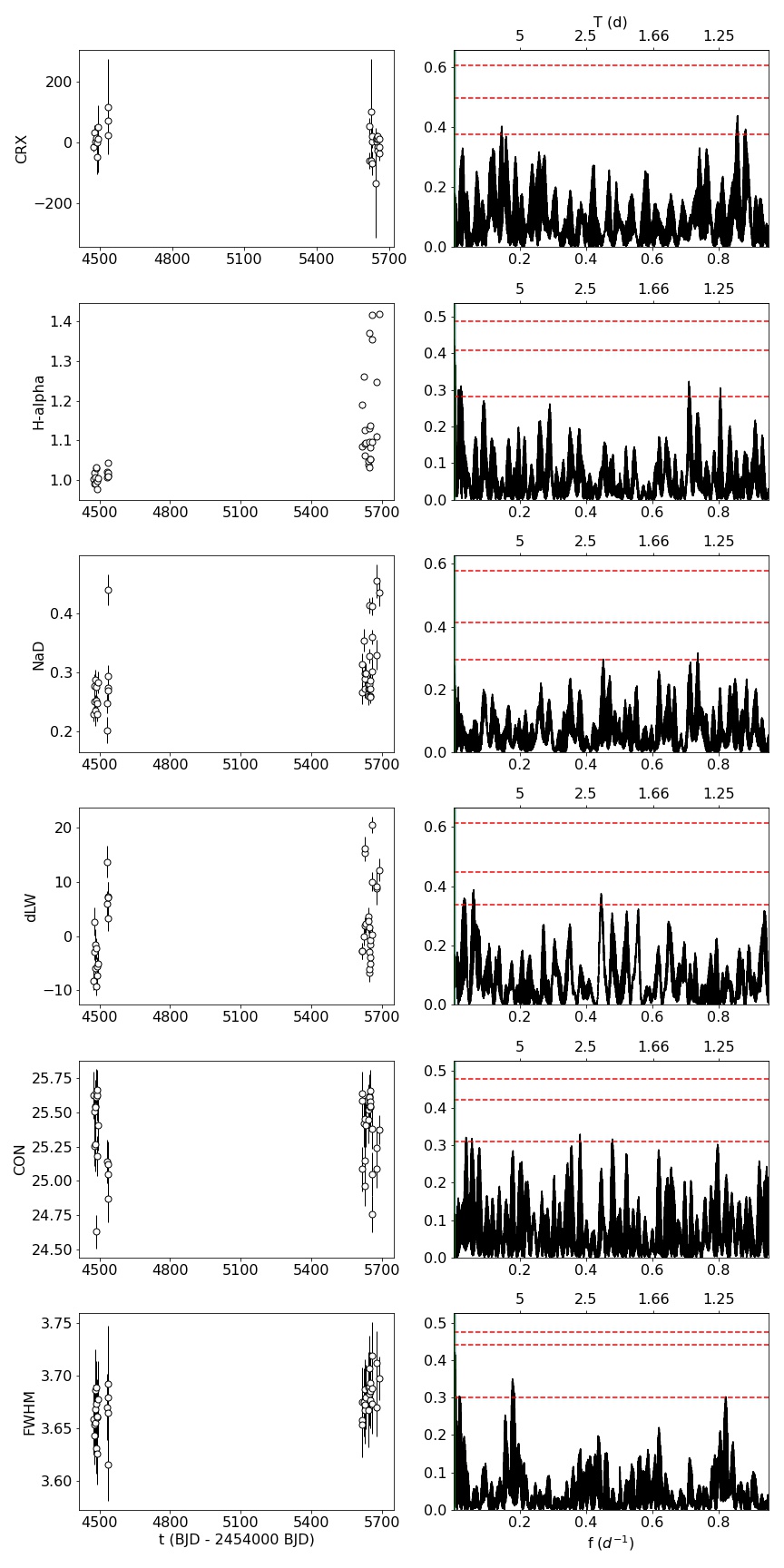}
 \caption{Spectroscopic index time-series data from HARPS-N, none of which show any significant signals. Since the periodograms are dominated by the window function, the low-frequency part is not shown. From top to bottom this is as indicated CRX, H$\alpha$, Na~{\sc i}~D, dLw, FWHM, and CON indices. Further details in Fig.\,\ref{prewhit1}.} \label{prewhit3}
\end{figure*}

The data and periodograms of the individual MEarth datasets are shown in Fig.\,\ref{prewhit1}, and those of additional activity indices from CARMENES and HARPS-N spectra are shown in Fig.\,\ref{prewhit2} and\,\ref{prewhit3}, respectively. 

\section{Corner plot of the best model}

\begin{figure*}
\centering
\includegraphics[width=\textwidth]{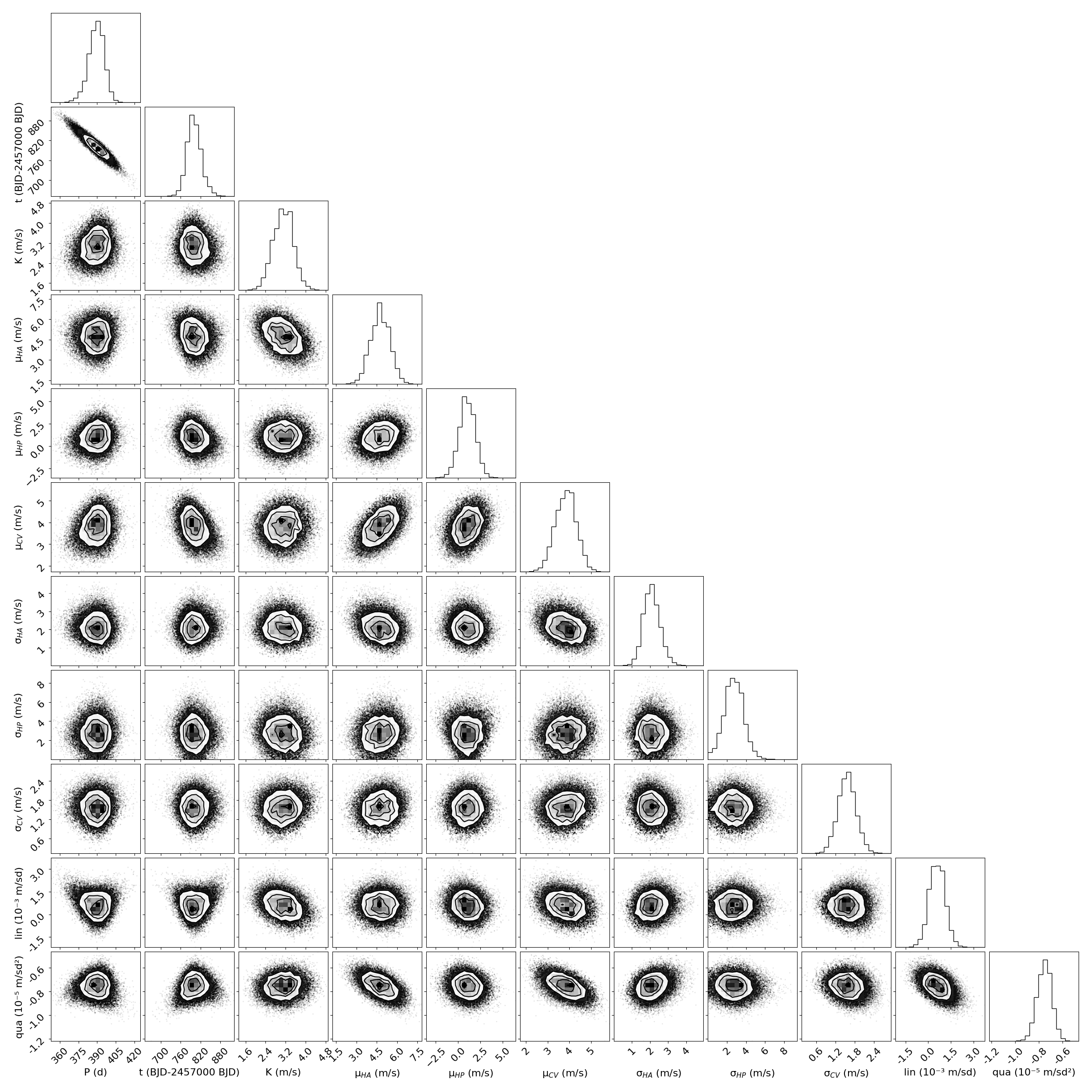}
 \caption{Posterior distribution of the fitted parameters of the sinusoid + parabola model.} \label{corner_sin_quad} 
\end{figure*}

The corner plot of the parameters of the best-fit model is given in Fig.\,\ref{corner_sin_quad}.

\end{appendix}

\end{document}